\documentclass[11p,reqno]{amsart}

\usepackage[T1]{fontenc}
\usepackage{graphicx}
\usepackage{amssymb,amsthm,amsmath,mathrsfs,bm,braket,marginnote}
\usepackage{enumerate}
\usepackage{appendix}
\usepackage[colorlinks=true, pdfstartview=FitV, linkcolor=blue, citecolor=blue, urlcolor=blue]{hyperref}
\usepackage{multirow}
\usepackage{pgf}
\usepackage{pgfplots}
\usepackage{tikz}
\usetikzlibrary{arrows,calc}
\usepackage{verbatim}
\usetikzlibrary{decorations.pathreplacing,decorations.pathmorphing}
\usepackage[numbers,sort&compress]{natbib}
\usepackage{dsfont}

\allowdisplaybreaks[1]
\numberwithin{equation}{section}
\newtheorem{theorem}{Theorem}[section]

\newtheorem{corollary}[theorem]{Corollary}

\newtheorem{proposition}[theorem]{Proposition}

 \reversemarginpar

\newcommand{\into}{\int_{(0,\infty)^2}}
\newcommand{\al}{{\alpha}}

\newcommand{\e}{{\rm e}}
\newcommand{\tr}{{\mathrm{tr}}}

\newcommand{\dd}{\,\mathrm{d}}

\begin{document}
\title[moments of quantum purity]{Moments of quantum purity and biorthogonal polynomial recurrence}
\author{Shi-Hao Li and Lu Wei}

\address{Shi-Hao Li, Department of Mathematics, Sichuan University, Chengdu, 610064, China}
\email{lishihao@lsec.cc.ac.cn;~shihao.li@scu.edu.cn}

\address{Lu Wei, Department of Electrical and Computer Engineering, University of Michigan, Dearborn, MI 48128, USA}
\email{luwe@umich.edu}

\subjclass[2010]{}
\date{\today}
\dedicatory{}
\maketitle

\begin{abstract}
The Bures-Hall ensemble is a unique measure of density matrices that satisfies various distinguished properties in quantum information processing. In this work, we study the statistical behavior of entanglement over the Bures-Hall ensemble as measured by the simplest form of an entanglement entropy - the quantum purity. The main results of this work are the exact second and third moment expressions of quantum purity valid for any subsystem dimensions, where the corresponding results in the literature are limited to the scenario of equal subsystem dimensions. In obtaining the results, we have derived recurrence relations of the underlying integrals over the Cauchy-Laguerre biorthogonal polynomials that may be of independent interest.
\end{abstract}

\section{Introduction and main results}
Quantum information theory aims at studying the theoretical foundations of quantum technologies including quantum computing and quantum communications. Crucial to successful exploitation of the quantum revolutionary advances is the understanding of the phenomenon of quantum entanglement. Entanglement is the most fundamental characteristic trait of quantum mechanics, which is also the resource that enables quantum technologies.

In this work, we study the statistical behavior of entanglement of quantum bipartite systems over the Bures-Hall ensemble~\cite{Hall98,Zyczkowski01,Sommers03}. In particular, we investigate the degree of entanglement as measured by quantum purity over such an ensemble. Quantum purity measures how far a state is from a pure state and is the simplest form of an entanglement entropy. We focus on finding the exact moments of purity, which in practice can be utilized to construct finite-size approximations to the distribution~\cite{Wei20,Wei20b}. The considered bipartite model is useful in describing the entanglement between the two subsystems of various real-world quantum systems, in which one subsystem represents a physical object (such as a set of spins) and the other subsystem is the environment (such as a heat bath). Existing results on the moments of purity are limited to the special case of equal subsystem dimensions, where the exact first three moments are known~\cite{Sommers04,Osipov10}. In the general case of subsystems of possibly unequal dimensions, the exact first moment formula has been recently obtained in~\cite{Sarkar19,Wei20a}. The contribution of this work lies in the corresponding exact second and third moment expressions of quantum purity.

The density matrix formalism~\cite{vN27} introduced by von Neumann provides a natural framework to describe density matrices of quantum states. Under this framework, the formulation that has led to the Bures-Hall ensemble~\cite{Hall98,Zyczkowski01,Sommers03} is outlined as follows.
Consider a bipartite system that consists of two subsystems $A$ and $B$ of Hilbert space (complex vector space) dimensions $m$ and $n$, respectively. The Hilbert space $\mathcal{H}_{A+B}$ of the composite system is given by the tensor product of the subsystems, $\mathcal{H}_{A+B}=\mathcal{H}_{A}\otimes\mathcal{H}_{B}$. A random pure state of the composite system $\mathcal{H}_{A+B}$ is defined as a linear combination of the coefficients $z_{i,j}$ and the complete bases $\left\{\Ket{i^{A}}\right\}$ and $\left\{\Ket{j^{B}}\right\}$ of $\mathcal{H}_{A}$ and $\mathcal{H}_{B}$,
\begin{equation}\label{eq:S0}
\Ket{\psi}=\sum_{i=1}^{m}\sum_{j=1}^{n}z_{i,j}\Ket{i^{A}}\otimes\Ket{j^{B}},
\end{equation}
where each $z_{i,j}$ follows the standard complex Gaussian distribution of zero mean and unit variance with the probability constraint $\sum_{i,j}|z_{i,j}|^2=1$. We now consider a superposition of the state~(\ref{eq:S0}) of the form
\begin{equation}\label{eq:SB}
\Ket{\varphi}=\Ket{\psi}+\left(\mathbf{U}\otimes\mathbf{I}_{m}\right)\Ket{\psi},
\end{equation}
where $\mathbf{U}$ is an $m\times m$ random unitary matrix with the measure proportional to $\det\left(\mathbf{I}_{m}+\mathbf{U}\right)^{2\alpha+1}$~\cite{Sarkar19}. The corresponding density matrix of the pure state~(\ref{eq:SB}) is
\begin{equation}\label{eq:rho}
\rho=\Ket{\varphi}\Bra{\varphi},
\end{equation}
which satisfies the natural probability constraint
\begin{equation}\label{eq:del}
\tr(\rho)=1.
\end{equation}
We assume without loss of generality that the dimension of subsystem $A$ is no greater than that of the subsystem $B$, i.e., $m\leq n$. The reduced density matrix $\rho_{A}$ of the smaller subsystem $A$ is computed by partial tracing (a.k.a. purification) of the full density matrix~(\ref{eq:rho}) over the other subsystem $B$, interpreted as the environment, as
\begin{equation}\label{eq:rhoB}
\rho_{A}=\tr_{B}\rho.
\end{equation}
The resulting density of eigenvalues of $\rho_{A}$ is the generalized\footnote{Hereafter, we refer to this generalized ensemble as the Bures-Hall ensemble despite the term is often referred to the special case $\alpha=-1/2$ in the literature.} Bures-Hall ensemble~\cite{Hall98,Zyczkowski01,Sommers03,Sarkar19}
\begin{equation}\label{eq:BH}
f\left(\bm{\lambda}\right)=\frac{1}{c}~\delta\left(1-\sum_{i=1}^{m}\lambda_{i}\right)\prod_{1\leq i<j\leq m}\frac{\left(\lambda_{i}-\lambda_{j}\right)^{2}}{\lambda_{i}+\lambda_{j}}\prod_{i=1}^{m}\lambda_{i}^{\alpha}
\end{equation}
supported in the probability simplex
\begin{equation}\label{eq:D}
\mathcal{D}=\Bigg\{0\leq\lambda_{m}<\ldots<\lambda_{1}\leq1,~~\sum_{i=1}^{m}\lambda_{i}=1\Bigg\},
\end{equation}
where the parameter $\alpha$ takes half-integer values
\begin{equation}\label{eq:aBH}
\alpha=n-m-\frac{1}{2}
\end{equation}
and the constant $c$ is
\begin{equation}\label{eq:cBH}
c=\frac{2^{-m(m+2\alpha)}\pi^{m/2}}{\Gamma\left(m(m+2\alpha+1)/2\right)}\prod_{i=1}^{m}\frac{\Gamma(i+1)\Gamma(i+2\alpha+1)}{\Gamma(i+\alpha+1/2)}.
\end{equation}
A relatively detailed derivation of the Bures-Hall ensemble~(\ref{eq:BH}) can be found, for example, in~\cite[Sec.~3]{Osipov10}. The presence of the Dirac delta function $\delta(\cdot)$ in~(\ref{eq:BH}) reflects the constraint~(\ref{eq:del}). Note that another approach to define the Bures-Hall ensemble~(\ref{eq:BH}) is by introducing a distance metric, known as the Bures-Hall metric, over reduced density matrices~\cite{BZ17}. The Bures-Hall ensemble satisfies several distinguished properties among the measures of random density matrices. It is the only monotone metric that is simultaneously Fisher adjusted and Fubini-Study adjusted. The Bures-Hall metric, related to quantum distinguishability, is known to be the minimal monotone metric~\cite{BZ17}. It is also a function of fidelity~\cite{Zski05}, a key performance indicator in quantum information processing. In addition, the Bures-Hall measure enjoys the property that, without any prior knowledge on a density matrix, the optimal way to estimate the density matrix is to generate a state at random with respect to this measure. It is therefore often used as a prior distribution, referred to as the Bures prior, in reconstructing quantum states from measurements.

Entanglement serves as a measure of the non-classical correlation between the subsystems $A$ and $B$. The degree of entanglement can be assessed by entanglement entropies, which are functions of the eigenvalues (entanglement spectrum) of a density matrix. Any function that satisfies a list of axioms qualifies as an entanglement entropy~\cite{BZ17}. In particular, an entropy should monotonically change from the separable state
\begin{equation}\label{eq:s}
\lambda_{1}=1,~~\lambda_{2}=\dots=\lambda_{m}=0
\end{equation}
to the maximally-entangled state
\begin{equation}\label{eq:e}
\lambda_{1}=\lambda_{2}=\dots\lambda_{m}=\frac{1}{m}
\end{equation}
corresponding to boundaries of the support~(\ref{eq:D}). In this work, we focus on the study of quantum purity
\begin{equation}\label{eq:P}
\text{P}=\sum_{i=1}^{m}\lambda_{i}^{2}
\end{equation}
supported in $\text{P}\in[1/m,1]$, which attains maximally-entangled state and separable state when $\text{P}=1/m$ and $\text{P}=1$, respectively. Quantum purity estimates how far a state is from a pure state that corresponds to $\text{P}=1$, i.e., when the density matrix becomes a one dimensional projector. Being a simple polynomial function, quantum purity is one of the few entanglement entropies that can be measured experimentally~\cite{Islam15}.

Statistical information of an entanglement entropy such as the considered quantum purity is encoded through its moments: the first moment (average value) implies the typical behavior of entanglement, the second moment (variance) specifies the fluctuation around the typical value, and the higher order moments (such as skewness and kurtosis) describe the tails of the distribution. Moreover, the moments can be also utilized to construct non-asymptotic approximations to the distribution of purity. Due to the compact support of purity, the resulting approximations become exact as the number of moments involved increases as promised by the Weierstrass approximation theorem. In the literature, the first moment of purity has been recently obtained in~\cite{Sarkar19,Wei20} as
\begin{equation}\label{eq:mP1}
\mathbb{E}_{f}\!\left[\text{P}\right]=\frac{m^2-2mn-4n^2-1}{2n\left(m^2-2mn-2\right)}.
\end{equation}
Closed-form expressions of the higher order moments remain open except for the special case of equal subsystem dimensions $m=n$ (i.e., $\alpha=-1/2$), where the exact second and third moments have been derived in~\cite{Osipov10} via an integrable system method. In this work, we propose a framework that allows one to obtain any higher order moment of purity of arbitrary subsystem dimensions in a systematic manner. A key ingredient is the recurrence relations of the underlying integrals of biorthogonal polynomials that we study in the first place. In particular, using the proposed framework we derive the second and third moments valid for any subsystem dimensions as summarized in the following proposition, which contains the main results of this work.
\begin{figure}[!h]
\centering
\includegraphics[width=0.74\linewidth]{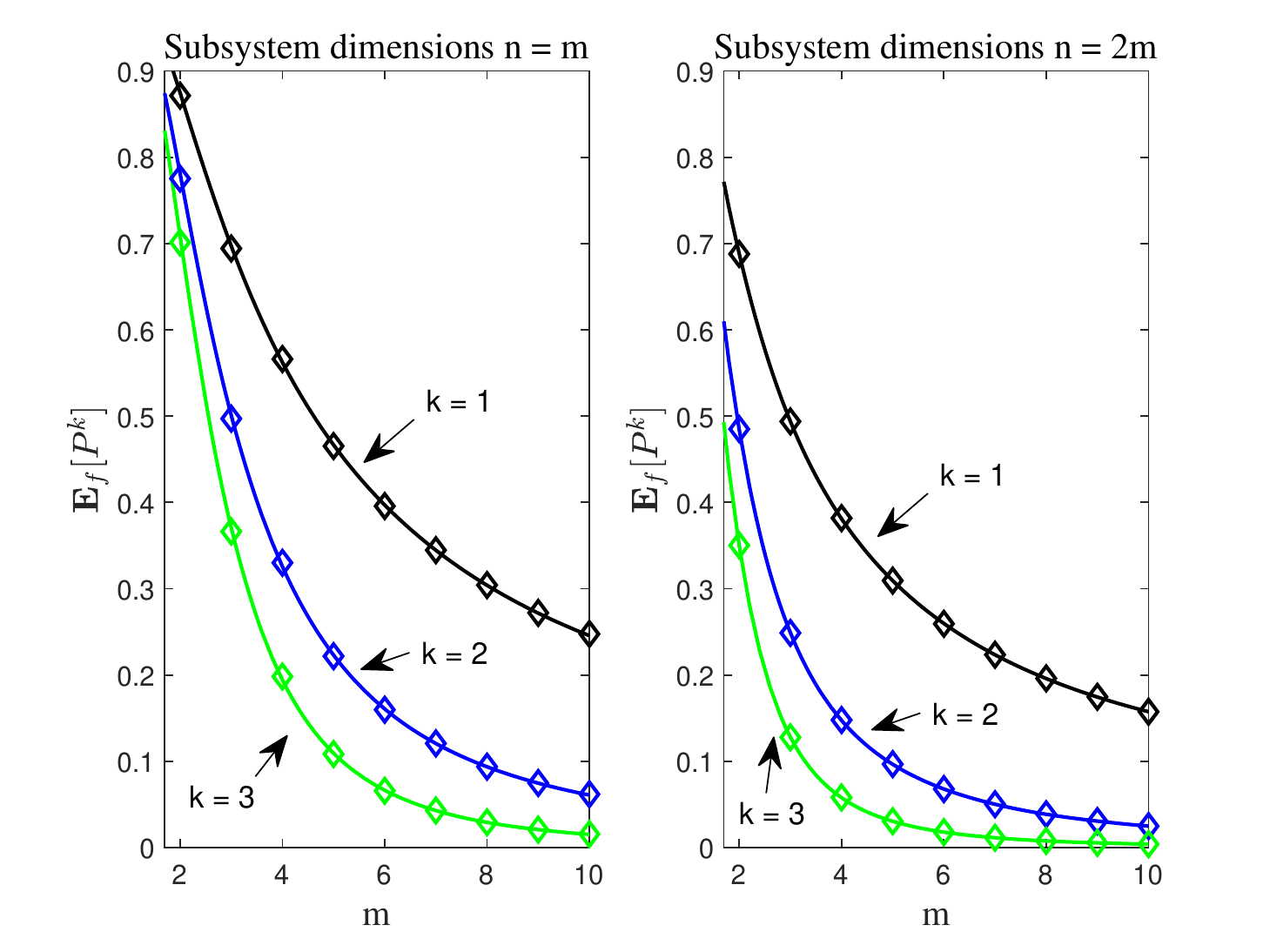}
\caption{The first three moments ($k=1,2,3$) of quantum purity $\mathbb{E}_{f}\!\left[\text{P}^{k}\right]$: analytical results versus simulations. Solid lines represent the obtained analytical results~(\ref{eq:mP1}),~(\ref{eq:mP2}),~(\ref{eq:mP3}), and the diamond-shape scatters represent numerical simulations. The left subplot and the right subplot refer to the cases of equal subsystem dimensions $n=m$ and unequal subsystem dimensions $n=2m$, respectively.}
\label{fig:p}
\end{figure}
\begin{proposition}\label{p:main}
For any subsystem dimensions $m\leq n$, the exact second moment and third moment of quantum purity~(\ref{eq:P}) over the Bures-Hall ensemble~(\ref{eq:BH}) are given respectively by
\begin{eqnarray}\label{eq:mP2}
\mathbb{E}_{f}\!\left[\text{P}^{2}\right]&=&\frac{1}{4\left(n^2-1\right)\prod_{i=1}^{3}\left(m^2-2mn-2i\right)}\big(m^6-6m^5n+4m^4n^2+6m^4+24m^3n^3-\nonumber\\
&&24m^3n-16m^2n^4+112m^2n^2-107m^2-32mn^5-176mn^3+214mn-\nonumber\\
&&128n^4-32n^2+160\big)
\end{eqnarray}
and
\begin{eqnarray}\label{eq:mP3}
\mathbb{E}_{f}\!\left[\text{P}^{3}\right]&=&\frac{1}{8n\left(n^2-1\right)\left(n^2-4\right)\prod_{i=1}^{5}\left(m^2-2mn-2i\right)}\big(m^{10}n^2-2m^{10}-10m^9n^3+20m^9n+\nonumber\\
&&28m^8n^4-35m^8n^2+22m^8+16m^7n^5-200m^7n^3-176m^7n-160m^6n^6+\nonumber\\
&&992m^6n^4-697m^6n^2+642m^6+64m^5n^7-1808m^5n^5+6646m^5n^3-3852m^5n+\nonumber\\
&&320m^4n^8+320m^4n^6-9208m^4n^4+13715m^4n^2-9262m^4-128m^3n^9+\nonumber\\
&&4480m^3n^7-10760m^3n^5-29180m^3n^3+37048m^3n-256m^2n^{10}-3520m^2n^8+\nonumber\\
&&30032m^2n^6-48052m^2n^4-6520m^2n^2+31640m^2-3072mn^9-18944mn^7+\nonumber\\
&&144192mn^5-61056mn^3-63280mn-10240n^8+15360n^6+132480n^4-\nonumber\\
&&114560n^2-23040\big).
\end{eqnarray}
\end{proposition}
\begin{proof} Detailed derivations of the results~(\ref{eq:mP2}) and~(\ref{eq:mP3}) are provided in the next section. \end{proof}
In the special case $m=n$, the results in proposition~\ref{p:main} are reduced respectively to
\begin{equation}\label{eq:P2mn}
\mathbb{E}_{f}\!\left[\text{P}^{2}\right]=\frac{5\left(5m^4+47m^2+32\right)}{4\left(m^2+2\right)\left(m^2+4\right)\left(m^2+6\right)}
\end{equation}
and
\begin{equation}\label{eq:P2mn}
\mathbb{E}_{f}\!\left[\text{P}^{3}\right]=\frac{5 \left(25 m^8+690 m^6+6015 m^4+8750 m^2+1152\right)}{8 m \left(m^2+2\right) \left(m^2+4\right) \left(m^2+6\right) \left(m^2+8\right) \left(m^2+10\right)}
\end{equation}
as reported\footnote{Notice the typo in the second moment expression in~\cite[equation~(60)]{Osipov10}, where the correct constant factor in the denominator is $4$ instead of $2$ as also verified by numerical simulations.} in~\cite[equation~(60)]{Osipov10}. Besides the finite-size results, we also point out that the asymptotic distribution of purity valid for large subsystem dimensions has been obtained in~\cite{Borot12} using a Coulomb gas method.

To illustrate the obtained results, we plot in figure~\ref{fig:p} the exact moments as compared with simulations, where the value of each scatter is computed by averaging over $10^{6}$ realizations of the Bures-Hall density matrices. We consider the scenarios of equal subsystem dimensions $n=m$ as well as unequal subsystem dimensions $n=2m$, where the claimed results match well with the simulations. By comparing the two subplots, we also observe that the values of the moments decrease as the dimension of the larger subsystem $n$ increases from $m$ to $2m$. The observation indicates that this additional subsystem size ($m$ to $2m$) leads to a system that tends to concentrate on more entangled states (i.e., with a smaller purity value), as expected.

The rest of the paper is organized as follows. In section~\ref{sec:2}, we provide detailed derivations of the main results on the exact second and third moments of quantum purity. Specifically, in section~\ref{sec:un} we relate the moment computation of any order to that over an unconstraint ensemble, whose correlation functions are given explicitly. In section~\ref{sec:m1}, we derive necessary results on the recurrence relations so as to provide a new proof to the recently obtained first moment formula. The second and third moments are derived respectively in section~\ref{sec:m2} and section~\ref{sec:m3}, where the recurrence relation framework is further developed. Potential future work is outlined in section~\ref{sec:con} after summarizing the main findings of the paper.

\section{Moment computation and orthogonal polynomial recurrence}\label{sec:2}
\subsection{Moment relation and correlation functions}\label{sec:un}
The first step is a rather standard procedure, see, e.g.,~\cite{Wei20,Wei20b,Osipov10,Sarkar19,Wei20a,Wei17,WW21}, of
relating the moment computation to that over an ensemble without the constraint $\delta\left(1-\sum_{i=1}^{m}\lambda_{i}\right)$ in~(\ref{eq:BH}). It turns out that the corresponding unconstraint ensemble is given by~\cite{Sarkar19,Wei20b,Wei20a}
\begin{equation}\label{eq:BHu}
h\left(\bm{x}\right)=\frac{1}{c'}\prod_{1\leq i<j\leq m}\frac{\left(x_{i}-x_{j}\right)^{2}}{x_{i}+x_{j}}\prod_{i=1}^{m}x_{i}^{\alpha}\e^{-x_{i}},
\end{equation}
where $x_{i}\in[0,\infty)$, $i=1,\dots,m$, and the constant $c'$ depends on the constant~(\ref{eq:cBH}) through
\begin{equation}\label{eq:cBHu}
c'=c~\Gamma\left(d\right)
\end{equation}
with $d$ denoting
\begin{equation}\label{eq:d}
d=\frac{1}{2}m\left(m+2\alpha+1\right).
\end{equation}
Beyond the physically relevant $\alpha$ values in~(\ref{eq:aBH}), the results hereafter are in fact valid for any $\alpha>-1$. It is known that the density $h\left(\bm{x}\right)$ in~\eqref{eq:BHu} admits the factorization~\cite{Wei20b}
\begin{equation}\label{eq:h2ft}
h(\bm{x})\prod_{i=1}^{m}\dd x_{i}=f(\bm{\lambda})g_{d}(\theta)\dd\theta\prod_{i=1}^{m}\dd\lambda_{i},
\end{equation}
where
\begin{equation}
g_{d}(\theta)=\frac{1}{\Gamma\left(d\right)}\e^{-\theta}\theta^{d-1}
\end{equation}
is the density of the trace of the unconstraint ensemble
\begin{equation}\label{eq:tr}
\theta=\sum_{i=1}^{m}x_{i},~~~~~~\theta\in[0,\infty).
\end{equation}
The factorization~(\ref{eq:h2ft}) implies that the random variable $\theta$ is independent of each of $\bm{\lambda}$.

Computing $k$-th moment of the purity $\text{P}$ can now be converted to computing $k$-th moment of the induced purity
\begin{equation}
\text{T}=\sum_{i=1}^{m}x_{i}^{2}
\end{equation}
over the unconstraint ensemble~(\ref{eq:BHu}) as
\begin{eqnarray}
\mathbb{E}_{f}\!\left[\text{P}^{k}\right]&=&\int_{\bm{\lambda}}\frac{\text{T}^{k}}{\theta^{2k}}f(\bm{\lambda})\prod_{i=1}^{m}\dd\lambda_{i} \\
&=&\int_{\bm{\lambda}}\frac{T^{k}}{\theta^{2k}}f(\bm{\lambda})\prod_{i=1}^{m}\dd\lambda_{i}\int_{\theta}g_{d+2k}(\theta)\dd\theta \\
&=&\frac{\Gamma(d)}{\Gamma(d+2k)}\int_{\bm{\lambda}}\text{T}^{k}f(\bm{\lambda})\prod_{i=1}^{m}\dd\lambda_{i}\int_{\theta}g_{d}(\theta)\dd\theta \\
&=&\frac{\Gamma(d)}{\Gamma(d+2k)}\mathbb{E}_{h}\!\left[\text{T}^{k}\right],\label{eq:P2T}
\end{eqnarray}
where we have used~(\ref{eq:h2ft}).

Now the task is to compute the induced moments $\mathbb{E}_{h}\!\left[\text{T}^{k}\right]$, the first three of which in terms of the first three point densities $h_{1}(x)$, $h_{2}(x,y)$, and $h_{3}(x,y,z)$ are written by definition as
\begin{subequations}\label{eq:T123}
\begin{eqnarray}
\mathbb{E}_{h}\!\left[\text{T}\right]&=&m\int_{0}^{\infty}x^{2}h_{1}(x)\dd x \label{eq:T1} \\
\mathbb{E}_{h}\!\left[\text{T}^{2}\right]&=&m\int_{0}^{\infty}\!\!x^{4}h_{1}(x)\dd x+m(m-1)\int_{0}^{\infty}\!\!\int_{0}^{\infty}\!\!x^{2}y^{2}~h_{2}\left(x,y\right)\dd x\dd y \label{eq:T2} \\
\mathbb{E}_{h}\!\left[\text{T}^{3}\right]&=&m\int_{0}^{\infty}\!\!x^{6}h_{1}(x)\dd x+3m(m-1)\int_{0}^{\infty}\!\!\int_{0}^{\infty}\!\!x^{4}y^{2}h_{2}(x,y)\dd x\dd y+ \nonumber\\
&&m(m-1)(m-2)\int_{0}^{\infty}\!\!\int_{0}^{\infty}\!\!\int_{0}^{\infty}\!\!x^{2}y^{2}z^{2}h_{3}(x,y,z)\dd x\dd y\dd z.\label{eq:T3}
\end{eqnarray}
\end{subequations}
In general, any $k$-point ($k\leq m$) density of the unconstraint ensemble is known to admit a Pfaffian form of a $2k\times2k$ antisymmetric matrix~\cite{forrester16}. In particular, the needed densities can be read off as functions of the four correlation kernels $K_{00}(x,y)$, $K_{01}(x,y)$, $K_{10}(x,y)$, and $K_{11}(x,y)$ as~\cite{forrester16}
\begin{eqnarray}
h_{1}(x)&=&\frac{1}{2m}\left(K_{01}(x,x)+K_{10}(x,x)\right)\label{eq:h1} \\
h_{2}(x,y)&=&\frac{1}{4m(m-1)}(\left(K_{01}(x,x)+K_{10}(x,x)\right)\left(K_{01}(y,y)+K_{10}(y,y)\right)-2K_{01}(x,y)K_{01}(y,x) \nonumber \\
&&-2K_{10}(x,y)K_{10}(y,x)-2K_{00}(x,y)K_{11}(x,y)-2K_{00}(y,x)K_{11}(y,x))\label{eq:h2} \\
h_{3}(x,y,z)&=&\frac{1}{8m(m-1)(m-2)}\left(h_{\text{A}}+h_{\text{B}}+h_{\text{C}}+h_{\text{D}}\right), \label{eq:h3}
\end{eqnarray}
where\begin{eqnarray}
h_{\text{A}} &=& (K_{01}(x,x)+K_{10}(x,x))(K_{01}(y,y)+K_{10}(y,y))(K_{01}(z,z)+K_{10}(z,z))\label{eq:hA}\\
h_{\text{B}} &=& -2(K_{01}(x,x)+K_{10}(x,x))(K_{01}(y,z)K_{01}(z,y)+K_{10}(y,z)K_{10}(z,y)+ \nonumber \\
&&K_{00}(y,z)K_{11}(y,z)+K_{00}(z,y)K_{11}(z,y))-2(K_{01}(y,y)+K_{10}(y,y))\times\nonumber \\
&&(K_{01}(x,z)K_{01}(z,x)+K_{10}(x,z)K_{10}(z,x)+K_{00}(x,z)K_{11}(x,z)+\nonumber \\
&&K_{00}(z,x)K_{11}(z,x))-2(K_{01}(z,z)+K_{10}(z,z))(K_{01}(x,y)K_{01}(y,x)+\nonumber \\
&&K_{10}(x,y)K_{10}(y,x)+K_{00}(x,y)K_{11}(x,y)+K_{00}(y,x)K_{11}(y,x))\label{eq:hB} \\
h_{\text{C}} &=& 2(K_{00}(x,y)K_{01}(y,z)K_{11}(x,z)+K_{00}(x,y)K_{10}(z,x)K_{11}(z,y)+ \nonumber \\
&&K_{00}(y,x)K_{01}(x,z)K_{11}(y,z)+K_{00}(y,x)K_{10}(z,y)K_{11}(z,x)+ \nonumber \\
&&K_{00}(x,z)K_{01}(z,y)K_{11}(x,y)+K_{00}(x,z)K_{10}(y,x)K_{11}(y,z)+ \nonumber \\
&&K_{00}(z,x)K_{01}(x,y)K_{11}(z,y)+K_{00}(z,x)K_{10}(y,z)K_{11}(y,x)+ \nonumber \\
&&K_{00}(y,z)K_{01}(y,x)K_{11}(x,z)+K_{00}(y,z)K_{10}(x,z)K_{11}(y,x)+ \nonumber \\
&&K_{00}(z,y)K_{01}(z,x)K_{11}(x,y)+K_{00}(z,y)K_{10}(x,y)K_{11}(z,x)- \nonumber \\
&&K_{00}(x,y)K_{01}(x,z)K_{11}(y,z)-K_{00}(x,y)K_{10}(z,y)K_{11}(z,x)- \nonumber \\
&&K_{00}(y,x)K_{01}(y,z)K_{11}(x,z)-K_{00}(y,x)K_{10}(z,x)K_{11}(z,y)- \nonumber \\
&&K_{00}(x,z)K_{01}(x,y)K_{11}(z,y)-K_{00}(x,z)K_{10}(y,z)K_{11}(y,x)- \nonumber \\
&&K_{00}(z,x)K_{01}(z,y)K_{11}(x,y)-K_{00}(z,x)K_{10}(y,x)K_{11}(y,z)- \nonumber \\
&&K_{00}(y,z)K_{01}(z,x)K_{11}(x,y)-K_{00}(y,z)K_{10}(x,y)K_{11}(z,x)- \nonumber \\
&&K_{00}(z,y)K_{01}(y,x)K_{11}(x,z)-K_{00}(z,y)K_{10}(x,z)K_{11}(y,x))\label{eq:hC} \\
h_{\text{D}} &=& 2(K_{01}(x,y)K_{01}(y,z)K_{01}(z,x)+K_{01}(x,z)K_{01}(z,y)K_{10}(x,y)+ \nonumber \\
&&K_{01}(x,z)K_{01}(y,x)K_{10}(y,z)+K_{01}(y,x)K_{01}(z,y)K_{10}(z,x)+ \nonumber \\
&&K_{01}(x,y)K_{10}(x,z)K_{10}(z,y)+K_{01}(y,z)K_{10}(x,z)K_{10}(y,x)+ \nonumber \\
&&K_{01}(z,x)K_{10}(y,x)K_{10}(z,y)+K_{10}(x,y)K_{10}(y,z)K_{10}(z,x)).\label{eq:hD}
\end{eqnarray}
To make the discussion self-contained, we provide a direct verification of the normalization constants of the above densities in the appendix~\ref{ap}. Note that the densities~(\ref{eq:h2}) and~(\ref{eq:h3}) have been simplified to the current form by using the following factorization properties\footnote{A proof of the factorization properties of the correlation kernels~(\ref{eq:vw}) can be found in~\cite{WW21}.}~\cite{forrester16}
\begin{subequations}\label{eq:vw}
\begin{eqnarray}
K_{00}(x,y)+K_{00}(y,x) &=& \ell_1(x)\ell_1(y) \label{eq:ww} \\
K_{01}(x,y)-K_{10}(y,x) &=& \ell_2(x)\ell_1(y) \label{eq:wv} \\
K_{11}(x,y)+K_{11}(y,x) &=& -\ell_2(x)\ell_2(y), \label{eq:vv}
\end{eqnarray}
\end{subequations}
where
\begin{subequations}\label{eq:fvw}
\begin{eqnarray}
\ell_1(x)&=&\sum_{i=0}^{m-1}\frac{\Gamma(i+m+2\alpha+2)(-x)^i}{\Gamma(i+2\alpha+2)\Gamma(i+\alpha+2)\Gamma(m-i)\Gamma(i+1)} \label{eq:w} \\
\ell_2(x)&=&-x^{2\alpha+1}\sum_{i=0}^{m-1}\frac{\Gamma(i+m+2\alpha+2)\Gamma(-i-\alpha,x)(-x)^{i}}{(i+\alpha+1)\Gamma(i+2\alpha+2)\Gamma(m-i)\Gamma(i+1)}+\e^{-x}x^{\alpha}\label{eq:v}
\end{eqnarray}
\end{subequations}
with $\Gamma(a,x)=\int_{x}^{\infty}t^{a-1}\e^{-t}\dd t$ denoting the incomplete Gamma function.

The correlation kernels admit two sets of representations useful in the subsequent calculations. The first one is the summation representation
\begin{subequations}\label{eq:ker}
\begin{eqnarray}
K_{00}(x,y)&=&\sum_{k=0}^{m-1}\frac{1}{h_k}p_{k}(x)q_{k}(y) \label{eq:K00} \\
K_{01}(x,y)&=&-x^{\alpha}e^{-x}\sum_{k=0}^{m-1}\frac{1}{h_k}p_{k}(y)Q_{k}(-x) \label{eq:K01} \\
K_{10}(x,y)&=&-y^{\alpha+1}e^{-y}\sum_{k=0}^{m-1}\frac{1}{h_k}P_{k}(-y)q_{k}(x) \label{eq:K10} \\
K_{11}(x,y)&=&x^{\alpha}y^{\alpha+1}e^{-x-y}\sum_{k=0}^{m-1}\frac{1}{h_k}P_{k}(-y)Q_{k}(-x)-W(x,y), \label{eq:K11}
\end{eqnarray}
\end{subequations}
where the weight function $W(x,y)$ of the (monic) Cauchy-Laguerre biorthogonal polynomials $p_{k}(x)$ and $q_{l}(y)$,
\begin{equation}\label{eq:oc}
\int_{0}^{\infty}\!\!\int_{0}^{\infty}p_{k}(x)q_{l}(y)W(x,y)\dd x\dd y=h_k\delta_{kl}
\end{equation}
is given by
\begin{equation}\label{eq:w}
W(x,y)=\frac{x^{\alpha}y^{\alpha+1}e^{-x-y}}{x+y}.
\end{equation}
The functions in~(\ref{eq:ker}) are further related by~\cite{bertola14,forrester16}
\begin{subequations}
\begin{eqnarray}\label{eq:PQ}
P_{k}(x)&=&\int_{0}^{\infty}\frac{v^{\alpha}\e^{-v}}{x-v}p_{k}(v)\dd v \\
Q_{k}(y)&=&\int_{0}^{\infty}\frac{w^{\alpha+1}\e^{-w}}{y-w}q_{k}(w)\dd w
\end{eqnarray}
\end{subequations}
known as the Cauchy transforms of $p_{k}(x)$ and $q_{k}(y)$, respectively. Moreover, they are expressed explicitly via Meijer G-functions as~\cite{BGS10,bertola14,forrester16}
\begin{subequations}\label{eq:kerM}
\begin{align}
p_{k}(x)&=(-1)^k \frac{\Gamma(k+1)\Gamma(2\alpha+k+2)\Gamma(\alpha+k+1)}{\Gamma(2\alpha+2k+2)}G_{2,3}^{1,1}\left(\left.
\begin{array}{c}
-2\alpha-k-1;~k+1\\
0;~-\alpha,\,-2\alpha-1\end{array}
\right|x
\right)\\
q_k(y)&=(-1)^k \frac{\Gamma(k+1)\Gamma(2\alpha+k+2)\Gamma(\alpha+k+2)}{\Gamma(2\alpha+2k+2)}G_{2,3}^{1,1}\left(\left.
\begin{array}{c}
-2\alpha-k-1;~k+1\\
0;-\alpha-1,\,-2\alpha-1
\end{array}
\right|y
\right)\\
P_k(x)&=(-1)^{k+1}\frac{2\alpha+2k}{\Gamma(k)\Gamma(\alpha+k)}G_{2,3}^{3,1}\left(
\left.\begin{array}{c}
-k;\,k+2\alpha\\
-1,\alpha-1,2\alpha;
\end{array}\right|-x
\right)\\
Q_k(y)&=(-1)^{k+1}\frac{2\alpha+2k}{\Gamma(k)\Gamma(\alpha+k+1)}G_{2,3}^{3,1}\left(
\left.\begin{array}{c}
-k;\,k+2\alpha\\
-1,\alpha,2\alpha;
\end{array}\right|-y
\right),
\end{align}
\end{subequations}
where the Meijer G-function is defined by the contour integral~\cite{PBM86}
\begin{eqnarray}\label{eq:MG}
&&G_{p,q}^{m,n}\left(\left.\begin{array}{c} a_{1},\ldots,a_{n}; a_{n+1},\ldots,a_{p} \\ b_{1},\ldots,b_{m}; b_{m+1},\ldots,b_{q} \end{array}\right|x.\right)\nonumber\\
=&&\frac{1}{2\pi\imath}\int_{\mathcal{L}}{\frac{\prod_{j=1}^m\Gamma\left(b_j+s\right)\prod_{j=1}^n\Gamma\left(1-a_j-s\right)x^{-s}}{\prod_{j=n+1}^p \Gamma\left(a_{j}+s\right)\prod_{j=m+1}^q\Gamma\left(1-b_j-s\right)}}\dd s
\end{eqnarray}
with the contour $\mathcal{L}$ separating the poles of $\Gamma\left(1-a_j-s\right)$ from the poles of $\Gamma\left(b_j+s\right)$. In addition to the summation representation~(\ref{eq:ker}), the correlation kernels also admit the integral form~\cite{bertola14}
\begin{subequations}\label{eq:kerI}
\begin{eqnarray}
K_{00}(x,y)&=&\int_{0}^{1}t^{2\alpha+1}H_{\alpha}(tx)H_{\alpha+1}(ty)\dd t \label{eq:K00I} \\
K_{01}(x,y)&=&x^{2\alpha+1}\int_{0}^{1}t^{2\alpha+1}H_{\alpha}(ty)G_{\alpha+1}(tx)\dd t \label{eq:K01I} \\
K_{10}(x,y)&=&y^{2\alpha+1}\int_{0}^{1}t^{2\alpha+1}H_{\alpha+1}(tx)G_{\alpha}(ty)\dd t \label{eq:K10I} \\
K_{11}(x,y)&=&(xy)^{2\alpha+1}\int_{0}^{1}t^{2\alpha+1}G_{\alpha+1}(tx)G_{\alpha}(ty)\dd t-\frac{x^{\alpha}y^{\alpha+1}}{x+y}, \label{eq:K11I}
\end{eqnarray}
\end{subequations}
where we denote
\begin{subequations}
\begin{eqnarray}
H_{q}(x)&=&G_{2,3}^{1,1}\left(\left.\begin{array}{c}-m-2\alpha-1;~m\\0;-q,-2\alpha-1\end{array}\right|x\Big.\right)\\
G_{q}(x)&=&G_{2,3}^{2,1}\left(\left.\begin{array}{c}-m-2\alpha-1;~m\\0,-q;-2\alpha-1\end{array}\right|x\Big.\right).
\end{eqnarray}
\end{subequations}

With the results introduced above, the moments~(\ref{eq:T123}) can now be computed by integrating over the appropriate form of the correlation kernels. However, the direct integration may not be an efficient method for the moment computation. We now discuss how the recurrence coefficients of the Cauchy-Laguerre biorthogonal polynomials are useful in obtaining the moments of purity. For positive integers $\beta$ and $\gamma$, the idea is to successively reduce the power of the average of $x^{\beta}y^{\gamma}$ over the ensemble to averages of lower order powers. The reduction is substantially simplified by the recurrence relations of the biorthogonal polynomials. Instead of performing integrations, the only task now is to collect non-zero contributions as identified by the biorthogonality condition. As a result, the desired integrals boil down to summations involving the recurrence coefficients of the biorthogonal polynomials, which are known explicitly. Besides the considered purity, the approach in fact works for any polynomial linear spectral statistics. To some extend, the idea is similar to that in Aomoto's proof of the Selberg integral~\cite[chapter~17.3-17.4]{Mehta}, which concerns the computation of averages over classical random matrix ensembles~\cite{Mehta,Forrester}. The proposed method also has independent interest in the field of combinatorics~\cite{hardy18} as briefly outlined here. Consider an oriented graph $G=(V,E)$ with vertices $V=\mathbb{N}^2$ and edges $E=\left\{(n,k)\to(n+1,m),\, n,k\in\mathbb{N},\, 0\leq m\leq k+1\right\}$, where the weight on the path is given by $\omega((n,k)\to(n+1,m))=\langle xp_k,q_m\rangle$. We then have
\begin{align*}
xp_k(x)=\sum_{\ell=0}^{k+1}\langle xp_k,q_\ell\rangle p_\ell(x),
\end{align*}
and this leads, by induction, to
\begin{align*}
x^n p_k(x)=\sum_{\ell=0}^{k+n}\left(
\sum_{\gamma:(0,k)\to(n,\ell)}\prod_{e\in\gamma}\omega(e)\right)p_\ell(x).
\end{align*}
The summation over the oriented paths $\gamma$ on $G$ starts from $(0,k)$ and ends at $(n,\ell)$, where each path picks the product of the weights along the edges it crosses.

\subsection{Computation of the first moment}\label{sec:m1}
To illustrate the proposed method, here we rederive the first moment of purity recently reported in~\cite{Sarkar19,Wei20a}. By the definition~(\ref{eq:T1}), one has
\begin{eqnarray}\label{eq:eht}
\mathbb{E}_{h}\!\left[\text{T}\right]=m\int_{0}^{\infty}\!\!x^{2}h_{1}(x)\dd x =\frac{1}{2}\int_{0}^{\infty}\!\!x^{2}K_{01}(x,x)\dd x+\frac{1}{2}\int_{0}^{\infty}\!\!x^{2}K_{10}(x,x)\dd x.
\end{eqnarray}
We now focus on the first integral
\begin{align}\label{eq:ter1}
\int_0^\infty\!\! x^2K_{01}(x,x)\dd x,
\end{align}
where the second one can be similarly computed. By the summation form of the kernels \eqref{eq:ker}, we have
\begin{eqnarray}\label{eq:P1}
\int_{0}^{\infty}\!\!x^{2}K_{01}(x,x)\dd x=\sum_{k=0}^{m-1}\frac{1}{h_k}\int_{0}^{\infty}\!\!\int_{0}^{\infty}\!\!x^{2}p_{k}(x)q_{k}(y)W(x,y)\dd x \dd y.
\end{eqnarray}
To compute the integral \eqref{eq:P1}, it is more convenient to consider a general form
\begin{align}\label{eq1}
\int_0^\infty\!\!\int_0^\infty\!\!x^\beta p_k(x)q_k(y)W(x,y)\dd x \dd y
\end{align}
with $\beta\in\mathbb{Z}_+$. To this end, we first state the following proposition.
\begin{proposition}
Let $\{a_{k,i}\}_{i\leq k}$ denote the coefficients of polynomials $p_k(x)$ in~(\ref{eq:K00}) for an arbitrary $k\in\mathbb{N}$, i.e.,
\begin{align*}
p_k(x)=x^k+a_{k,k-1}x^{k-1}+\cdots+a_{k,0}x^0
\end{align*}
with the convention $a_{k,k}=1$. For the Cauchy-Laguerre biorthogonal polynomials defined by \eqref{eq:oc}-\eqref{eq:w}, one has
\begin{align}
\int_0^\infty\!\!\int_0^\infty\!\!x^\beta q_k(y)W(x,y)\dd x \dd y=b_{\beta,\beta-k}h_k,\quad \beta\geq k,
\end{align}
where $h_k$ is the normalization constant in \eqref{eq:oc} and $b_{\beta,k}$ is recursively given by
\begin{align}\label{rbni}
b_{\beta,i}=-\sum_{j=0}^{i-1}b_{\beta,j}a_{\beta-j,\beta-i},\quad b_{\beta,0}=1.
\end{align}
\end{proposition}
\begin{proof}
The idea is to write $x^\beta$ in terms of the orthogonal basis $\{p_k(x)\}_{k\leq \beta}$ before applying the orthogonality condition. When $\beta=k$, it is obvious that
\begin{align*}
\int_0^\infty\!\!\int_0^\infty x^kq_k(y)W(x,y)\dd x\dd y=h_k
\end{align*}
by orthogonality. When $\beta=k+1$, since
\begin{align*}
x^{k+1}=p_{k+1}(x)-a_{k+1,k}p_k(x)+\text{l.o.t.},
\end{align*}
the lower order terms ($\text{l.o.t.}$) do not contribute to the result when taking the inner product with $q_k(y)$ on both sides, one finds that
\begin{align*}
\int_0^\infty\!\!\int_0^\infty x^{k+1}q_k(y)W(x,y)\dd x\dd y=-a_{k+1,k}h_k.
\end{align*}
Continuing this procedure establishes the result of this proposition.
\end{proof}
Moreover, if we expand
\begin{align*}
x^\beta p_k(x)=x^{\beta+k}+a_{k,k-1}x^{\beta+k-1}+\cdots+a_{k,0}x^\beta
\end{align*}
and employ the result of above proposition, we then have
\begin{align}\label{eq:o1}
&\frac{1}{h_k}\int_0^\infty\!\!\int_0^\infty x^\beta p_k(x)q_k(y)W(x,y)\dd x\dd y=
\sum_{j=0}^\beta a_{k,k-j}b_{\beta+k-j,\beta-j}.
\end{align}
Similarly, if we expand $q_k(y)$ as
\begin{align*}
q_k(y)=y^{k}+\hat{a}_{k,k-1}y^{k-1}+\cdots+\hat{a}_{k,0}y^0,
\end{align*}
then by the recursion
\begin{align}\label{eq:rhbni}
\hat{b}_{\beta,i}=-\sum_{j=0}^{i-1}\hat{b}_{\beta,j}\hat{a}_{\beta-j,\beta-i},\quad \hat{b}_{\beta,0}=1,
\end{align}
one has
\begin{align*}
\frac{1}{h_k}\int_0^\infty\!\!\int_0^\infty y^\beta p_k(x)q_k(y)W(x,y)dxdy=\sum_{j=0}^\beta \hat{a}_{k,k-j}\hat{b}_{\beta+k-j,\beta-j}.
\end{align*}
Putting the results together, one has
\begin{align}\label{eq:ehtalt}
\mathbb{E}_h[\text{T}]=\frac{1}{2}\sum_{k=0}^{m-1}\sum_{j=0}^2\left(
a_{k,k-j}b_{k+2-j,2-j}+\hat{a}_{k,k-j}\hat{b}_{k+2-j,2-j}
\right),
\end{align}
where the explicit formulas for $\{a_{k,j},\, \hat{a}_{k,j}\}_{j\leq k}$ are given by~\cite[equation~(2.6)]{forrester16}
\begin{align}\label{eq:ea}
\begin{aligned}
a_{k,j}&=(-1)^{k-j}{k\choose j}\frac{\Gamma(2a+k+j+2)\Gamma(2a+k+2)\Gamma(a+k+1)}{\Gamma(2a+2k+2)\Gamma(2a+j+2)\Gamma(a+j+1)}\\
\hat{a}_{k,j}&=(-1)^{k-j}{k\choose j}\frac{\Gamma(2a+k+j+2)\Gamma(2a+k+2)\Gamma(a+k+2)}{\Gamma(2a+2k+2)\Gamma(2a+j+2)\Gamma(a+j+2)}.
\end{aligned}
\end{align}
Inserting the above $a_{k,j}$ and $\hat{a}_{k,j}$ into \eqref{rbni} and \eqref{eq:rhbni}, respectively, we claim that
\begin{align}\label{bnj}
\begin{aligned}
b_{k,j}&={k\choose j}\frac{\Gamma(2a+k+2)\Gamma(a+k+1)\Gamma(2a+2k+3-2j)}{\Gamma(2a+k+2-j)\Gamma(a+k+1-j)\Gamma(2a+2k+3-j)}\\
\hat{b}_{k,j}&={k\choose j}\frac{\Gamma(2a+k+2)\Gamma(a+k+2)\Gamma(2a+2k+3-2j)}{\Gamma(2a+k+2-j)\Gamma(a+k+2-j)\Gamma(2a+2k+3-j)},
\end{aligned}
\end{align}
which is equivalent of stating that
\begin{align}\label{eq:bCON}
\sum_{j=0}^{i-1}(-1)^j {i \choose j}\frac{(2\al+2k+2-2j)\Gamma(2\al+2k-i-j+2)}{\Gamma(2\al+2k+3-j)}=(-1)^{i-1}\frac{\Gamma(2\al+2k+3-2i)}{\Gamma(2\al+2k+3-i)}.
\end{align}
The left hand side of the above identity is equal to
\begin{eqnarray}
&&\sum_{j=0}^{i-1}\frac{(-1)^{j}\Gamma(i+1)}{\Gamma(j+1)\Gamma(i+1-j)}\frac{(2\al+2k+2-2j)\Gamma(2\al+2k-i-j+2)}{\Gamma(2\al+2k+3-j)} \nonumber\\
&=&(-1)^{i-1}\sum_{\ell=0}^{i-1}\frac{(-1)^{\ell}\Gamma(i+1)}{\Gamma(\ell+2)\Gamma(i-\ell)}\frac{(2\al+2k+4-2i+2\ell)\Gamma(2\al+2k+3-2i+\ell)}{\Gamma(2\al+2k+4-i+\ell)}\nonumber\\
&=&\frac{(-1)^i}{\Gamma(-i)}\sum_{\ell=0}^{i-1}\frac{(2\al+2k+4-2i+2\ell)\Gamma(2\al+2k+3-2i+\ell)\Gamma(1-i+\ell)}{\Gamma(2\al+2k+4-i+\ell)\Gamma(2+\ell)}\nonumber\\
&=&\frac{(-1)^i}{\Gamma(-i)}\frac{1}{-i}\left(\frac{\Gamma(2\al+2k+3-i)}{\Gamma(2\al+2k+3)\Gamma(i+1)}-\frac{\Gamma(1-i)\Gamma(2\al+2k-2i+3)}{\Gamma(2\al+2k-i+3)}\right)\nonumber\\
&=&(-1)^{i-1}\frac{\Gamma(2\al+2k-2i+3)}{\Gamma(2\al+2k-i+3)},
\end{eqnarray}
where the second to last equality is a direct application of \cite[lemma 4.1]{bertola14}.

Finally, inserting~(\ref{eq:ea}) and~(\ref{bnj}) into \eqref{eq:ehtalt}, one obtains
\begin{equation}
\mathbb{E}_{h}\!\left[\text{T}\right]=\frac{m(2\alpha+m+1)\left(4\alpha^2+4\alpha+5m^2+10\alpha m+5m+2\right)}{4(2\alpha+2m+1)},
\end{equation}
which upon applying the moment relation~(\ref{eq:P2T}) for $k=1$ gives
\begin{eqnarray}
\mathbb{E}_{f}\!\left[\text{P}\right]&=&\frac{4}{m(m+2\alpha+1)\left(m^2+2\alpha m+m+2\right)}\mathbb{E}_{h}\!\left[\text{T}\right]\nonumber\\
&=&\frac{5m^2+10\alpha m+5m+4\alpha^2+4\alpha+2}{(2m+2\alpha+1)\left(m^2+2\alpha m+m+2\right)}.
\end{eqnarray}
This verifies the mean purity formula~(\ref{eq:mP1}) obtained in~\cite{Sarkar19,Wei20a} when replacing $\alpha$ by~(\ref{eq:aBH}).

\subsection{Computation of the second moment}\label{sec:m2}
We now proceed to the computation of the second moment $\mathbb{E}_{h}\!\left[\text{T}^2\right]$. Recall that
\begin{align*}
\mathbb{E}_{h}\!\left[\text{T}^{2}\right]=m\int_{0}^{\infty}\!\!x^{4}h_{1}(x)\dd x+m(m-1)\int_{0}^{\infty}\!\!\int_{0}^{\infty}\!\!x^{2}y^{2}~h_{2}\left(x,y\right)\dd x\dd y,
\end{align*}
where the first term is equal to
\begin{align*}
\frac{1}{2}\int_0^\infty x^4 (K_{01}(x,x)+K_{10}(x,x))\dd x=\frac{1}{2}\sum_{k=0}^{m-1}\sum_{j=0}^4\left(
a_{k,k-j}b_{k+4-j,4-j}+\hat{a}_{k,k-j}\hat{b}_{k+4-j,4-j}\right).
\end{align*}
The second term becomes $\mathbb{E}_{h}\!\left[\text{T}^{2}\right]-\mathcal{J}$ if we denote
\begin{eqnarray}\label{eq:j}
\mathcal{J}&=&\frac{1}{2}\int_{0}^{\infty}\!\!\int_{0}^{\infty}\!\!x^2y^2
\left(K_{01}(x,y)K_{01}(y,x)+K_{10}(x,y)K_{10}(y,x)\right)\dd x\dd y-\nonumber\\
&&\int_{0}^{\infty}\!\!\int_{0}^{\infty}\!\!x^2y^2K_{00}(x,y)K_{11}(x,y)\dd x\dd y.
\end{eqnarray}
The identity
\begin{eqnarray*}
&&\int_{0}^{\infty}\!\!\int_{0}^{\infty}\!\!x^2y^2 K_{01}(x,y)K_{01}(y,x)\dd x \dd y \\
&=&\sum_{j,k=0}^{m-1}\left(\frac{1}{h_j}\int_{0}^{\infty}\!\!\int_{0}^{\infty}\!\!x^2p_k(x)q_j(y)W(x,y)\dd x \dd y
\right)\left(\frac{1}{h_k}\int_{0}^{\infty}\!\!\int_{0}^{\infty}\!\! x^2p_j(x)q_k(y)W(x,y)\dd x \dd y\right)
\end{eqnarray*}
inspires us to consider a more general case
\begin{align*}
\frac{1}{h_j}\int_{0}^{\infty}\!\!\int_{0}^{\infty}\!\! x^\beta p_k(x)q_j(y)W(x,y)\dd x\dd y,\quad \beta\in\mathbb{N}
\end{align*}
for later use.
\begin{corollary}
For arbitrary $\beta,\, k,\, j\in\mathbb{N}$, we have
\begin{align*}
\frac{1}{h_j}\int_{0}^{\infty}\!\!\int_{0}^{\infty}\!\! x^\beta p_k(x)q_j(y)W(x,y)\dd x\dd y&=
\sum_{\ell=0}^{k}a_{k,\ell}b_{\beta+\ell,\beta+\ell-j}\\
\frac{1}{h_j}\int_{0}^{\infty}\!\!\int_{0}^{\infty}\!\! y^\beta p_j(x)q_k(y)W(x,y)\dd x\dd y&=\sum_{\ell=0}^{k}\hat{a}_{k,\ell}\hat{b}_{\beta+\ell,\beta+\ell-j}.
\end{align*}
\end{corollary}
This corollary is a generalization of the result~\eqref{eq:o1} and it should be remarked that when $\beta+\ell<j$, the term $b_{\beta+\ell,\beta+\ell-j}$ is zero. Therefore, noting that $\{p_k(x)\}_{k\in\mathbb{N}}$ and $\{q_k(y)\}_{k\in\mathbb{N}}$ are dual to each other, we have
\begin{eqnarray}
&&\int_{0}^{\infty}\!\!\int_{0}^{\infty}\!\! x^2y^2 K_{01}(x,y)K_{01}(y,x)\dd x\dd y \nonumber\\
&=&\sum_{j,k=0}^{m-1}\left(\sum_{\ell=0}^k a_{k,\ell}b_{\ell+2,\ell+2-j}\right)\left(\sum_{\ell=0}^j a_{j,\ell}b_{\ell+2,\ell+2-k}\right)\label{eq234}
\end{eqnarray}
and
\begin{eqnarray}
&&\int_{0}^{\infty}\!\!\int_{0}^{\infty}\!\! x^2y^2 K_{10}(x,y)K_{10}(y,x)\dd x\dd y \nonumber\\
&=&\sum_{j,k=0}^{m-1}\left(\sum_{\ell=0}^k \hat{a}_{k,\ell}\hat{b}_{\ell+2,\ell+2-j}\right)\left(
\sum_{\ell=0}^j \hat{a}_{j,\ell}\hat{b}_{\ell+2,\ell+2-k}\right).\label{eq234b}
\end{eqnarray}
We proceed to evaluate~\eqref{eq234}. By changing the order of summation, one obtains
\begin{align*}
\sum_{\ell_1=0}^{m-1}\sum_{\ell_2=0}^{m-1}\left(
\sum_{j=\ell_1}^{m-1}a_{j,\ell_2}b_{\ell_1+2,\ell_1+2-j}
\right)\left(
\sum_{k=\ell_2}^{m-1}a_{k,\ell_1}b_{\ell_2+2,\ell_2+2-k}
\right),
\end{align*}
and by substituting \eqref{eq:ea} and \eqref{bnj} into above equation, it leads to
\begin{eqnarray*}
\sum_{j=\ell_1}^{m-1}a_{j,\ell_2}b_{\ell_1+2,\ell+2-j}&=&(-1)^{m+\ell_1}\frac{\Gamma(\ell_2+3)\Gamma(2\al+\ell_2+4)\Gamma(\al+\ell_2+3)}{\Gamma(\ell_1+1)\Gamma(2\al+\ell_1+2)\Gamma(\al+\ell_1+1)}\times\\
&&\frac{(m-\ell_1)(4+2\al+m+\ell_2)\Gamma(2a+\ell_1+m+2)}{(\ell_1-\ell_2-2)\Gamma(m-\ell_1+1)\Gamma(\ell_2-m+3)\Gamma(2\al+m+\ell_2+5)}.
\end{eqnarray*}
Therefore, the summation \eqref{eq234} is evaluated to
\begin{align*}
&-\frac{m^2(2\al+m+1)^2}{48(2\al+2m+1)^2}\big(8\al^2-56\al^4+12\al m-28\al^2m-264\al^3m+m^2-66\al m^2-454\al^2m^2+\\
&8\al^4m^2-32m^3-324\al m^3+4\al^2m^3+24\al^3m^3-81m^4+6\al m^4+26\al^2m^4+2m^5+12\al m^5+2m^6\big),
\end{align*}
where the one~(\ref{eq234b}) can be similarly obtained. For the last integral in~(\ref{eq:j}),
\begin{align}
\int_{0}^{\infty}\!\!\int_{0}^{\infty}\!\! x^2y^2K_{00}(x,y)K_{11}(x,y)\dd x\dd y,
\end{align}
we proceed to study a generalized one
\begin{equation}\label{eq:K00K11}
\int_{0}^{\infty}\!\!\int_{0}^{\infty}\!\! x^{\beta_{1}}y^{\beta_{2}}K_{00}(x,y)K_{11}(x,y)\dd x \dd y
\end{equation}
with $\beta_{1}$ and $\beta_{2}$ being non-negative integers. By using the summation form of the kernels~(\ref{eq:ker}) and the explicit polynomial expressions of $p_{k}(x)$ and $q_{k}(y)$, the resulting integrals over $x$ and $y$ can be separately evaluated by the Mellin transform of Meijer G-function~\cite{PBM86}
\begin{eqnarray}\label{eq:iMG}
&&\int_{0}^{\infty}\!\!x^{s-1}G_{p,q}^{m,n}\left(\left.\begin{array}{c} a_{1},\ldots,a_{n}; a_{n+1},\ldots,a_{p} \\ b_{1},\ldots,b_{m}; b_{m+1},\ldots,b_{q} \end{array}\right|\eta x\Big.\right)\dd x\nonumber\\
&=&\frac{\eta^{-s}\prod_{j=1}^m\Gamma\left(b_j+s\right)\prod_{j=1}^{n}\Gamma\left(1-a_j-s\right)}{\prod_{j=n+1}^{p}\Gamma\left(a_{j}+s\right)\prod_{j=m+1}^q\Gamma\left(1-b_j-s\right)}.
\end{eqnarray}
This leads to
\begin{eqnarray}
&&\int_{0}^{\infty}\!\!\int_{0}^{\infty}\!\! x^{\beta_{1}}y^{\beta_{2}}K_{00}(x,y)K_{11}(x,y)\dd x\dd y \label{eq:di}\\
&=&\sum_{j=0}^{m-1}\sum_{k=0}^{m-1}\sum_{i=0}^{j}\sum_{s=0}^{j}\frac{4(-1)^{i+s}(\alpha+j+1)(\alpha+k+1)\Gamma(2\alpha+i+j+2)\Gamma(2\alpha+j+s+2)}{i!s!\Gamma(\alpha+i+1)\Gamma(2\alpha+i+2)\Gamma(\alpha+s+2)\Gamma(2\alpha+s+2)}\times\nonumber\\
&&\frac{\Gamma\left(\beta_1+i+1\right)\Gamma\left(\alpha+\beta_1+i+1\right)\Gamma\left(2\alpha+\beta_1+i+2\right)}{\Gamma\left(\beta_1+i-k+1\right)\Gamma\left(2\alpha+\beta_1+i+k+3\right)\Gamma(j-i+1)}\times\nonumber\\
&&\frac{\Gamma\left(\beta_2+s+1\right)\Gamma\left(\alpha+\beta_2+s+2\right)\Gamma\left(2\alpha+\beta_2+s+2\right)}{\Gamma\left(\beta_2-k+s+1\right)\Gamma\left(2\alpha+\beta_2+k+s+3\right)\Gamma(j-s+1)}-\sum_{j=0}^{m-1}\sum_{i=0}^{j}\sum_{k=0}^{j}\frac{2(-1)^{i+k}}{i!k!}\times\nonumber\\
&&\frac{(\alpha+j+1)\Gamma(2\alpha+i+j+2)\Gamma(2\alpha+j+k+2)}{\Gamma(j-i+1)\Gamma(j-k+1)\Gamma(\alpha+i+1)\Gamma(2\alpha+i+2)\Gamma(\alpha+k+2)\Gamma(2\alpha+k+2)}\times\nonumber\\
&&\frac{\Gamma\left(\alpha+\beta_1+i+1\right)\Gamma\left(\alpha+\beta_2+k+2\right)}{2\alpha+\beta_1+\beta_2+i+k+2}\label{eq:4sums}\\
&=&\sum_{i=m-\beta_1}^{m-1}\sum_{k=m-\beta_2}^{m-1}\frac{(-1)^{i+k-1}\Gamma(2\alpha+i+m+2)\Gamma(2\alpha+k+m+2)}{(2\alpha+i+k+2)\left(2\alpha+\beta_1+\beta_2+i+k+2\right)i!k!\Gamma(m-i)\Gamma(m-k)}\times\nonumber\\
&&\frac{\Gamma\left(\beta_1+i+1\right)\Gamma\left(\beta_2+k+1\right)\Gamma\left(\alpha+\beta_1+i+1\right)\Gamma\left(\alpha+\beta_2+k+2\right)}{\Gamma(\alpha+i+1)\Gamma(2\alpha+i+2)\Gamma(\alpha+k+2)\Gamma(2\alpha+k+2)}\times\nonumber\\
&&\frac{\Gamma\left(2\alpha+\beta_1+i+2\right)\Gamma\left(2\alpha+\beta_2+k+2\right)}{\Gamma\left(\beta_1+i-m+1\right)\Gamma\left(\beta_2+k-m+1\right)\Gamma\left(2\alpha+\beta_1+i+m+2\right)\Gamma\left(2\alpha+\beta_2+k+m+2\right)},\nonumber
\end{eqnarray}
where the summations over $j$ and $k$ in~(\ref{eq:4sums}) have been evaluated by the identities, cf.~\cite[equations~(71)-(72)]{Wei20b},
\begin{equation}
\sum_{j=i}^{m-1}\frac{(\alpha+j+1)\Gamma(2\alpha+i+j+2)\Gamma(2\alpha+j+s+2)}{\Gamma(j-i+1)\Gamma(j-s+1)}=\frac{\Gamma(i+m+2\alpha+2)\Gamma(s+m+2\alpha+2)}{2(2\alpha+i+s+2)\Gamma(m-i)\Gamma(m-s)},\\
\end{equation}
\begin{eqnarray*}
&&\!\!\!\!\!\!\sum_{k=0}^{m-1}\frac{(\alpha+k+1)\left(\Gamma\left(\beta_1+i-k+1\right)\Gamma\left(\beta_2-k+s+1\right)\right)^{-1}}{\Gamma\left(2\alpha+\beta_1+i+k+3\right)\Gamma\left(2\alpha+\beta_2+k+s+3\right)}=\frac{1}{2\left(2\alpha+\beta_1+\beta_2+i+s+2\right)}\times\nonumber\\
&&\!\!\!\!\!\!\left(\frac{\left(\Gamma\left(i+\beta_1+1\right)\Gamma\left(s+\beta_2+1\right)\right)^{-1}}{\Gamma\left(i+2\alpha+\beta_1+2\right)\Gamma\left(s+2\alpha+\beta_2+2\right)}-\frac{\left(\Gamma\left(i-m+\beta_1+1\right)\Gamma\left(s-m+\beta_2+1\right)\right)^{-1}}{\Gamma\left(i+m+2\alpha+\beta_1+2\right)\Gamma\left(s+m+2\alpha+\beta_2+2\right)}\right).
\end{eqnarray*}
In the interested case $\beta_{1}=\beta_{2}=2$, one readily obtains
\begin{eqnarray}\label{eq:242}
&&\into x^{2}y^{2}K_{00}(x,y)K_{11}(x,y)\dd x\dd y \\
&=&-\frac{m^{2}(2\alpha+m+1)^2}{8(2\alpha+2m-1)(2\alpha+2m+1)^{2}(2\alpha+2m+3)}\big(32\alpha^4+64\alpha^3+12\alpha^2-20\alpha+33m^4+\nonumber\\
&&132\alpha m^3+66m^3+196\alpha^{2}m^2+196\alpha m^2+15m^2+128\alpha^{3}m+192\alpha^{2}m+28\alpha m-18m-6\big).\nonumber
\end{eqnarray}

Combining results of~\eqref{eq234}, \eqref{eq234b}, and \eqref{eq:242}, we finally arrive at
\begin{eqnarray}
\mathbb{E}_h\!\left[\text{T}^2\right]&=&\frac{1}{2}\sum_{k=0}^{m-1}\sum_{j=0}^{4}\left(a_{k,k-j}b_{k+4-j,4-j}+\hat{a}_{k,k-j}\hat{b}_{k+4-j,4-j}\right)+\mathbb{E}_{h}^{2}\!\left[\text{T}\right]-\mathcal{J}\\
&=&\frac{m(2\alpha+m+1)}{16(2\alpha+2m-1)(2\alpha+2m+3)}\big(128\alpha^4+256\alpha^3+224\alpha^2+96\alpha+25m^6+\nonumber\\
&&150\alpha m^5+75m^5+340\alpha^2m^4+340\alpha m^4+295m^4+360\alpha^3m^3+540\alpha^2m^3+\nonumber\\
&&1110\alpha m^3+465m^3+176\alpha^4m^2+352\alpha^3m^2+1448\alpha^2m^2+1272\alpha m^2+232m^2+\nonumber\\
&&32\alpha^5 m+80\alpha^4m+768\alpha^3m+1072\alpha^2m+376\alpha m+12m-144\big).\label{eq:T2}
\end{eqnarray}
Employing the moment relation~(\ref{eq:P2T}) for $k=2$ gives the second moment of quantum purity
\begin{equation}\label{eq:P2}
\mathbb{E}_{f}\!\left[\text{P}^{2}\right]=\frac{\Gamma(d)}{\Gamma(d+4)}\mathbb{E}_{h}\!\left[\text{T}^{2}\right],
\end{equation}
which upon inserting $\alpha=n-m-1/2$ leads to the claimed result~(\ref{eq:mP2}).

\subsection{Computation of the third moment}\label{sec:m3}
As shown in~(\ref{eq:T3}), computing the third moment boils down to computing the following three integrals
\begin{eqnarray}
\text{I}_{1}&=&\int_{0}^{\infty}\!\!x^{6}h_{1}(x)\dd x \\
\text{I}_{2}&=&\int_{0}^{\infty}\!\!\int_{0}^{\infty}\!\!x^{4}y^{2}h_{2}(x,y)\dd x\dd y \\
\text{I}_{3}&=&\int_{0}^{\infty}\!\!\int_{0}^{\infty}\!\!\int_{0}^{\infty}\!\!x^{2}y^{2}z^{2}h_{3}(x,y,z)\dd x\dd y\dd z,\label{eq:I3}
\end{eqnarray}
where the first two integrals can be obtained in the same manner as in the first two moments calculation. They are given by
\begin{eqnarray}
\text{I}_{1}&=&\frac{1}{2}\sum_{k=0}^{m-1}\sum_{j=0}^6\left(a_{k,k-j}b_{k+6-j,6-j}+\hat{a}_{k,k-j}\hat{b}_{k+6-j,6-j}\right)\\
&=&\frac{1}{32\prod_{i=1}^{5}(2\alpha+2m+2i-5)}\big(1024 \alpha ^{11}+5632 \alpha ^{10}+48640 \alpha ^9+176640 \alpha ^8+278592 \alpha ^7+\nonumber\\
&&190176 \alpha ^6-522880 \alpha ^5-1398640 \alpha ^4-930016 \alpha ^3-10608 \alpha ^2+260640 \alpha +7293 m^{11}+\nonumber\\
&&87516 \alpha  m^{10}+43758 m^{10}+463320 \alpha ^2 m^9+463320 \alpha  m^9+139425 m^9+1424280 \alpha ^3 m^8+\nonumber\\
&&2136420 \alpha ^2 m^8+1304160 \alpha  m^8+296010 m^8+2814240 \alpha ^4 m^7+5628480 \alpha ^3 m^7+\nonumber\\
&&5268120 \alpha ^2 m^7+2453880 \alpha  m^7+253539 m^7+3734016 \alpha ^5 m^6+9335040 \alpha ^4 m^6+\nonumber\\
&&12046320 \alpha ^3 m^6+8734440 \alpha ^2 m^6+1728012 \alpha  m^6-280566 m^6+3370752 \alpha ^6 m^5+\nonumber\\
&&10112256 \alpha ^5 m^5+17149440 \alpha ^4 m^5+17445120 \alpha ^3 m^5+5120808 \alpha ^2 m^5-1916376 \alpha m^5-\nonumber\\
&&1049565 m^5+2048640 \alpha ^7 m^4+7170240 \alpha ^6 m^4+15708000 \alpha ^5 m^4+21344400 \alpha ^4 m^4+\nonumber\\
&&8657880 \alpha ^3 m^4-4772460 \alpha ^2 m^4-5481960 \alpha  m^4-1404810 m^4+808320 \alpha ^8 m^3+\nonumber\\
&&3233280 \alpha ^7 m^3+9199680 \alpha ^6 m^3+16282560 \alpha ^5 m^3+9014040 \alpha ^4 m^3-5337360 \alpha ^3 m^3-\nonumber\\
&&10695600 \alpha ^2 m^3-5852040 \alpha  m^3-806532 m^3+192000 \alpha ^9 m^2+864000 \alpha ^8 m^2+\nonumber\\
&&3285120 \alpha ^7 m^2+7465920 \alpha ^6 m^2+5693184 \alpha ^5 m^2-2415840 \alpha ^4 m^2-9654240 \alpha ^3 m^2-\nonumber\\
&&8908560 \alpha ^2 m^2-2411328 \alpha  m^2+49608 m^2+23552 \alpha ^{10} m+117760 \alpha ^9 m+635520 \alpha ^8 m+\nonumber\\
&&1835520 \alpha ^7 m+1970304 \alpha ^6 m-18816 \alpha ^5 m-3921560 \alpha ^4 m-5905840 \alpha ^3 m-\nonumber\\
&&2513736 \alpha ^2 m+165456 \alpha  m+246240 m+86400\big)
\end{eqnarray}
and
\begin{eqnarray}
\text{I}_{2}&=&\frac{(2\alpha+m)(2\alpha+m+1)}{64(2\alpha+2m-3)(2\alpha+2m-1)(2\alpha+2m+1)(2\alpha+2m+5)}\big(256 \alpha ^8-1152 \alpha ^7+\nonumber\\
&&4032 \alpha ^6-1440 \alpha ^5-22656 \alpha ^4+55152 \alpha ^3-60832 \alpha ^2+33840 \alpha +1155 m^8+9240 \alpha  m^7-\nonumber\\
&&1518 m^7+31164 \alpha ^2 m^6-11802 \alpha  m^6+1506 m^6+57624 \alpha ^3 m^5-37896 \alpha ^2 m^5+\nonumber\\
&&10074 \alpha  m^5+6024 m^5+63504 \alpha ^4 m^4-64992 \alpha ^3 m^4+30456 \alpha ^2 m^4+32502 \alpha  m^4-\nonumber\\
&&32145 m^4+42336 \alpha ^5 m^3-63936 \alpha ^4 m^3+51168 \alpha ^3 m^3+60576 \alpha ^2 m^3-137754 \alpha  m^3+\nonumber\\
&&52158 m^3+16448 \alpha ^6 m^2-35712 \alpha ^5 m^2+47744 \alpha ^4 m^2+44688 \alpha ^3 m^2-203716 \alpha ^2 m^2+\nonumber\\
&&174276 \alpha  m^2-51156 m^2+3328 \alpha ^7 m-10304 \alpha ^6 m+22432 \alpha ^5 m+8944 \alpha ^4 m-\nonumber\\
&&119864 \alpha ^3 m+179416 \alpha ^2 m-116424 \alpha  m+28296 m-7200\big).
\end{eqnarray}
The $\text{I}_{3}$ integral is divided into four parts
\begin{equation}\label{eq}
\text{I}_{3}=A+B+C+D
\end{equation}
that correspond to integrals over the blocks of terms of~(\ref{eq:hA}),~(\ref{eq:hB}),~(\ref{eq:hC}), and~(\ref{eq:hD}), respectively. The integrals $A$ and $B$ are readily available based on the results of the first two moments and are given by
\begin{eqnarray}
A&=&\frac{m^3 (2 \alpha +m+1)^3 \left(4 \alpha ^2+4 \alpha +5 m^2+10 \alpha  m+5 m+2\right)^3}{64 m (m-1) (m-2) (2 \alpha +2 m+1)^3}\\
B&=&-\frac{3 m (2 \alpha +m+1)^2 \left(4 \alpha ^2+4 \alpha +5 m^2+10 \alpha  m+5 m+2\right)}{64 (m-2) (m-1) (2 \alpha +2 m-1) (2 \alpha +2 m+1)^3 (2 \alpha +2 m+3)}\big(128 \alpha ^7-64 \alpha ^6-\nonumber\\
&&160 \alpha ^5+272 \alpha ^4-160 \alpha ^3-64 \alpha ^2+48 \alpha +462 m^7+3234 \alpha  m^6+717 m^6+9324 \alpha ^2 m^5+\nonumber\\
&&3924 \alpha  m^5+219 m^5+14280 \alpha ^3 m^4+8316 \alpha ^2 m^4+546 \alpha  m^4+243 m^4+12432 \alpha ^4 m^3+\nonumber\\
&&8448 \alpha ^3 m^3+72 \alpha ^2 m^3+1008 \alpha  m^3-117 m^3+6048 \alpha ^5 m^2+3984 \alpha ^4 m^2-768 \alpha ^3 m^2+\nonumber\\
&&1632 \alpha ^2 m^2-336 \alpha  m^2-96 m^2+1472 \alpha ^6 m+576 \alpha ^5 m-688 \alpha ^4 m+1152 \alpha ^3 m-\nonumber\\
&&328 \alpha ^2 m-168 \alpha  m+12 m\big).
\end{eqnarray}
By the symmetry, the integral $C$ is written as the following eight distinct integrals
\begin{equation}\label{eq:C}
C=\frac{1}{2m(m-1)(m-2)}\left(2C_{1}+C_{2}-2C_{3}-C_{4}+2C_{5}+C_{6}-2C_{7}-C_{8}\right),
\end{equation}
where
\begin{eqnarray}
C_{1} &=& \int_{0}^{\infty}\!\!\int_{0}^{\infty}\!\!\int_{0}^{\infty}\!\!x^{2}y^{2}z^{2}K_{00}(x,y)K_{01}(y,z)K_{11}(x,z)\dd x\dd y\dd z\\
C_{2} &=& \int_{0}^{\infty}\!\!\int_{0}^{\infty}\!\!\int_{0}^{\infty}\!\!x^{2}y^{2}z^{2}K_{00}(y,z)K_{01}(y,x)K_{11}(x,z)\dd x\dd y\dd z\\
C_{3} &=& \int_{0}^{\infty}\!\!\int_{0}^{\infty}\!\!\int_{0}^{\infty}\!\!x^{2}y^{2}z^{2}K_{00}(x,y)K_{01}(x,z)K_{11}(y,z)\dd x\dd y\dd z\\
C_{4} &=& \int_{0}^{\infty}\!\!\int_{0}^{\infty}\!\!\int_{0}^{\infty}\!\!x^{2}y^{2}z^{2}K_{00}(y,z)K_{01}(z,x)K_{11}(x,y)\dd x\dd y\dd z\\
C_{5} &=& \int_{0}^{\infty}\!\!\int_{0}^{\infty}\!\!\int_{0}^{\infty}\!\!x^{2}y^{2}z^{2}K_{00}(x,y)K_{10}(z,x)K_{11}(z,y)\dd x\dd y\dd z\\
C_{6} &=& \int_{0}^{\infty}\!\!\int_{0}^{\infty}\!\!\int_{0}^{\infty}\!\!x^{2}y^{2}z^{2}K_{00}(y,z)K_{10}(x,z)K_{11}(y,x)\dd x\dd y\dd z\\
C_{7} &=& \int_{0}^{\infty}\!\!\int_{0}^{\infty}\!\!\int_{0}^{\infty}\!\!x^{2}y^{2}z^{2}K_{00}(x,y)K_{10}(z,y)K_{11}(z,x)\dd x\dd y\dd z\\
C_{8} &=& \int_{0}^{\infty}\!\!\int_{0}^{\infty}\!\!\int_{0}^{\infty}\!\!x^{2}y^{2}z^{2}K_{00}(y,z)K_{10}(x,y)K_{11}(z,x)\dd x\dd y\dd z.
\end{eqnarray}
Similarly, the integral $D$ is divided into
\begin{equation}\label{eq:D}
D=\frac{1}{4m(m-1)(m-2)}\left(D_{1}+3D_{2}+3D_{3}+D_{4}\right),
\end{equation}
where
\begin{eqnarray}
D_{1} &=& \int_{0}^{\infty}\!\!\int_{0}^{\infty}\!\!\int_{0}^{\infty}\!\!x^{2}y^{2}z^{2}K_{01}(x,y)K_{01}(y,z)K_{01}(z,x)\dd x\dd y\dd z \\
D_{2} &=& \int_{0}^{\infty}\!\!\int_{0}^{\infty}\!\!\int_{0}^{\infty}\!\!x^{2}y^{2}z^{2}K_{01}(x,z)K_{01}(z,y)K_{10}(x,y)\dd x\dd y\dd z \\
D_{3} &=& \int_{0}^{\infty}\!\!\int_{0}^{\infty}\!\!\int_{0}^{\infty}\!\!x^{2}y^{2}z^{2}K_{01}(x,y)K_{10}(x,z)K_{10}(z,y)\dd x\dd y\dd z \\
D_{4} &=& \int_{0}^{\infty}\!\!\int_{0}^{\infty}\!\!\int_{0}^{\infty}\!\!x^{2}y^{2}z^{2}K_{10}(x,y)K_{10}(y,z)K_{10}(z,x)\dd x\dd y\dd z.
\end{eqnarray}

The integrals in $C$ and $D$ can be written in terms of biorthogonal polynomials and the obtained recurrence relations play an important role in the evaluation of these integrals. On the other hand, they can also be directly calculated by integrations over the correlation kernels as illustrated in the calculation of~(\ref{eq:di}). In doing so, it will be efficient to utilize the integral representation~(\ref{eq:K01I}) and~(\ref{eq:K10I}) for the kernels $K_{01}(x,y)$ and $K_{10}(x,y)$ and the summation form~(\ref{eq:K00}) and~(\ref{eq:K11}) for the kernels $K_{00}(x,y)$ and $K_{11}(x,y)$, respectively. The resulting integrals are evaluated by using the identity~\cite{PBM86}
\begin{eqnarray}
&&\int_{0}^{1}\!x^{a-1}G_{p,q}^{m,n}\left(\left.\begin{array}{c} a_{1},\ldots,a_{n}; a_{n+1},\ldots,a_{p} \\ b_{1},\ldots,b_{m}; b_{m+1},\ldots,b_{q} \end{array}\right|\eta x\Big.\right)\dd x\nonumber\\ &=&G_{p+1,q+1}^{m,n+1}\left(\left.\begin{array}{c}1-a,a_{1},\ldots,a_{n};a_{n+1},\ldots,a_{p}\\b_{1},\ldots,b_{m};b_{m+1},\ldots,b_{q},-a\end{array}\right|\eta\Big.\right)
\end{eqnarray}
as well as the the Mellin transform of Meijer G-function~(\ref{eq:iMG}). The results are given by
\begin{equation}
C=-\frac{3m(m+1)(2\alpha+m+1)^{2}\gamma}{8(m-2)(2\alpha+2m-3)(2\alpha+2m-1)(2\alpha+2m+1)^{3}(2\alpha+2m+3)}
\end{equation}
and
\begin{equation}
D=\frac{\delta}{16(m-2)(m-1)(2\alpha+2m-3)(2\alpha+2m-1)(2\alpha+2m+1)^3},
\end{equation}
where $\gamma$ and $\delta$ are respectively
\begin{eqnarray}
\gamma&=&32\alpha^6-96\alpha^5-40\alpha^4+120\alpha^3+8\alpha^2-24\alpha+15m^6+120\alpha m^5+384\alpha^2m^4-36\alpha m^4-\nonumber \\
&&48m^4+624\alpha^3m^3-192 \alpha^2m^3-234\alpha m^3+27m^3+536\alpha^4m^2-368\alpha^3m^2-386\alpha^2m^2+\nonumber \\
&&128\alpha m^2+224\alpha^5m-304\alpha^4m-240\alpha^3m+208\alpha^2m+4 \alpha m-12m
\end{eqnarray}
and
\begin{eqnarray}
\delta&=&1024\alpha^{11}-9728\alpha^{10}+36352\alpha^9-64512\alpha^8+41280\alpha^7+30624\alpha^6-51136\alpha^5+1712\alpha^4+\nonumber \\
&&17696\alpha^3-1296\alpha ^2-2016\alpha+7293m^{11}+87516\alpha m^{10}-12726m^{10}+463320\alpha^2m^9-\nonumber \\
&&158004\alpha m^9+10539 m^9+1424280\alpha^3 m^8-844524\alpha^2m^8+140976\alpha m^8-5268m^8+\nonumber \\
&&2814240\alpha^4m^7-2560296\alpha^3m^7+778452\alpha^2m^7-78504\alpha m^7-20997m^7+3734016\alpha^5m^6-\nonumber \\
&&4871040\alpha^4m^6+2355888\alpha^3m^6-441336\alpha^2m^6-140292\alpha m^6+22338 m^6+3370752\alpha^6m^5-\nonumber \\
&&6061440\alpha^5m^5+4330752\alpha^4m^5-1295856\alpha^3m^5-365136\alpha^2 m^5+164100\alpha m^5-5331m^5+\nonumber \\
&&2048640\alpha^7m^4-4972224\alpha^6m^4+5023200\alpha^5m^4-2226960\alpha^4 m^4-438984\alpha^3m^4+\nonumber \\
&&485940\alpha^2m^4-47952\alpha m^4-8424 m^4+808320\alpha^8 m^3-2632704\alpha^7 m^3+3668544\alpha^6m^3-\nonumber \\
&&2317824\alpha^5m^3-169896\alpha^4m^3+737688\alpha^3m^3-159732\alpha^2 m^3-37032\alpha m^3+6048m^3+\nonumber \\
&&192000\alpha^9m^2-848640\alpha^8m^2+1615488\alpha^7m^2-1430208\alpha^6 m^2+134400\alpha^5m^2+\nonumber \\
&&594912\alpha^4m^2-248640\alpha^3m^2-52704\alpha^2m^2+25704\alpha m^2+624 m^2+23552\alpha^{10}m-\nonumber \\
&&146432\alpha^9m+383616\alpha^8m-476160\alpha^7m+152448\alpha^6m+232512\alpha^5m-182552\alpha^4m-\nonumber \\
&&24064\alpha^3m+36216\alpha^2m+576\alpha m-1008m.
\end{eqnarray}
Putting everything together, one obtains the third moment formula as
\begin{eqnarray}
\mathbb{E}_{f}\!\left[\text{P}^{3}\right]&=&\frac{1}{(2\alpha+m+1)\prod_{i=1}^{5}(2\alpha+2m+2i-5)(m^2+2\alpha m+m+2i)}\big(81920 \alpha ^9+\nonumber \\&&368640 \alpha ^8+614400 \alpha ^7+430080 \alpha ^6-1059840 \alpha ^5-2864640 \alpha ^4-1894400 \alpha ^3-\nonumber \\
&&7680 \alpha ^2+529920 \alpha +500 m^{13}+7000 \alpha  m^{12}+3500 m^{12}+43700 \alpha ^2 m^{11}+\nonumber \\
&&43700 \alpha  m^{11}+22225 m^{11}+160400 \alpha ^3 m^{10}+240600 \alpha ^2 m^{10}+255900 \alpha  m^{10}+\nonumber \\
&&87850 m^{10}+384160 \alpha ^4 m^9+768320 \alpha ^3 m^9+1294380 \alpha ^2 m^9+910220 \alpha  m^9+\nonumber \\
&&256845 m^9+629600 \alpha ^5 m^8+1574000 \alpha ^4 m^8+3783400 \alpha ^3 m^8+4101100 \alpha ^2 m^8+\nonumber \\
&&2386800 \alpha  m^8+562350 m^8+719616 \alpha ^6 m^7+2158848 \alpha ^5 m^7+7068480 \alpha ^4 m^7+\nonumber \\
&&10538880 \alpha ^3 m^7+9593004 \alpha ^2 m^7+4683372 \alpha  m^7+503935 m^7+572928 \alpha ^7 m^6+\nonumber \\
&&2005248 \alpha ^6 m^6+8807232 \alpha ^5 m^6+17004960 \alpha ^4 m^6+21867552 \alpha ^3 m^6+\nonumber \\
&&16798992 \alpha ^2 m^6+3491468 \alpha  m^6-535010 m^6+311040 \alpha ^8 m^5+1244160 \alpha ^7 m^5+\nonumber \\
&&7396992 \alpha ^6 m^5+17836416 \alpha ^5 m^5+31093200 \alpha ^4 m^5+33910560 \alpha ^3 m^5+\nonumber \\
&&10570948 \alpha ^2 m^5-3618956 \alpha  m^5-2077165 m^5+109568 \alpha ^9 m^4+493056 \alpha ^8 m^4+\nonumber \\
&&4122624 \alpha ^7 m^4+12128256 \alpha ^6 m^4+28475808 \alpha ^5 m^4+42019344 \alpha ^4 m^4+\nonumber \\
&&18278216 \alpha ^3 m^4-8866596 \alpha ^2 m^4-10838736 \alpha  m^4-2813290 m^4+22528 \alpha ^{10} m^3+\nonumber \\
&&112640 \alpha ^9 m^3+1452288 \alpha ^8 m^3+5133312 \alpha ^7 m^3+16659456 \alpha ^6 m^3+32484864 \alpha ^5 m^3+\nonumber \\
&&19378192 \alpha ^4 m^3-9621472 \alpha ^3 m^3-21137424 \alpha ^2 m^3-11734304 \alpha  m^3-1622340 m^3+\nonumber \\
&&2048 \alpha ^{11} m^2+11264 \alpha ^{10} m^2+289792 \alpha ^9 m^2+1219584 \alpha ^8 m^2+5914752 \alpha ^7 m^2+\nonumber \\
&&15089088 \alpha ^6 m^2+12371840 \alpha ^5 m^2-4003744 \alpha ^4 m^2-19106720 \alpha ^3 m^2-\nonumber \\
&&17907952 \alpha ^2 m^2-4858272 \alpha  m^2+102600 m^2+24576 \alpha ^{10} m+122880 \alpha ^9 m+\nonumber \\
&&1124352 \alpha ^8 m+3760128 \alpha ^7 m+4305408 \alpha ^6 m+271872 \alpha ^5 m-7811712 \alpha ^4 m-\nonumber \\
&&11935488 \alpha ^3 m-5078784 \alpha ^2 m+344448 \alpha  m+496800 m+172800\big),
\end{eqnarray}
which upon inserting $\alpha=n-m-1/2$ leads to the claimed result~(\ref{eq:mP3}). Finally, we note that the higher order moments beyond the first three can also be obtained in a similar manner but with an increased effort. Another potential approach to obtain the exact moments is via the method of integrable systems, where, for example, it was established in~\cite{Hu17} that the partition function of Bures-Hall ensemble is the $\tau$-function of certain integrable hierarchies.

\section{Conclusion and outlook}\label{sec:con}
In this work, we derived the exact second and third moments of quantum purity of arbitrary subsystem dimensions over the Bures-Hall ensemble, generalizing the results of equal subsystem dimensions in the literature. The derivations are obtained based on the direct integrations over the correlation kernels and the established recurrence relations of integrals involving biorthogonal polynomials. Future work includes devising an integrable system approach to systematically produce moments of any order as well as the study of asymptotic distribution of purity of large subsystem dimensions.

\section*{Acknowledgments}
We thank Santosh Kumar for providing the simulation codes and Jiyuan Zhang for correspondence. The work of Lu Wei is supported in part by the U.S. National Science Foundation (CNS-2006612).

\appendix
\section{Normalizations of Bures-Hall $k$-point densities}\label{ap}
In this appendix, we verify the normalization constants of the first three point densities~(\ref{eq:h1}),~(\ref{eq:h2}), and~(\ref{eq:h3}) of the Bures-Hall ensemble, which in particular makes use of the fact that the kernels $K_{01}(x,y)$, $K_{10}(x,y)$, and $K_{11}(x,y)$ can be written in terms of the kernel $K_{00}(x,y)$ as~\cite{forrester16}
\begin{subequations}\label{eq:kerI}
\begin{eqnarray}
K_{01}(x,y)&=&x^{\alpha}e^{-x}\int_{0}^{\infty}\frac{v^{\alpha+1}\e^{-v}}{x+v}K_{00}(y,v)\dd v\label{eq:K01II}\\
K_{10}(x,y)&=&y^{\alpha+1}e^{-y}\int_{0}^{\infty}\frac{w^{\alpha}\e^{-w}}{y+w}K_{00}(w,x)\dd w\label{eq:K10II} \\
K_{11}(x,y)&=&x^{\alpha}y^{\alpha+1}e^{-x-y}\int_{0}^{\infty}\!\!\int_{0}^{\infty}\frac{v^{\alpha}\e^{-v}}{y+v}\frac{w^{\alpha+1}\e^{-w}}{x+w}K_{00}(v,w)\dd v\dd w-W(x,y).\label{eq:K11II}
\end{eqnarray}
\end{subequations}
We first note the normalization of $K_{00}(x,y)$,
\begin{equation}
\int_{0}^{\infty}\!\!\int_{0}^{\infty}\!\!K_{00}(x,y)W(x,y)\dd x\dd y=\sum_{k=0}^{m-1}1=m
\end{equation}
and the reproducing properties of the correlation kernels
\begin{eqnarray}
\int_{0}^{\infty}\!\!\int_{0}^{\infty}\!\!K_{00}(x,z)K_{00}(w,y)W(w,z)\dd w\dd z&=&K_{00}(x,y)\\
\int_{0}^{\infty}K_{01}(x,y)K_{01}(y,z)\dd y&=&K_{01}(x,z)\label{eq:re01}\\
\int_{0}^{\infty}K_{10}(x,y)K_{10}(y,z)\dd y&=&K_{10}(x,z).\label{eq:re10}
\end{eqnarray}

For the one-point density~(\ref{eq:h1}), the normalization constant $2m$ is obtained as
\begin{eqnarray}
\int_{0}^{\infty}K_{01}(x,x)\dd x&=&\int_{0}^{\infty}\!\!\int_{0}^{\infty}x^{\alpha}e^{-x}\frac{v^{\alpha+1}\e^{-v}}{x+v}K_{00}(x,v)\dd v\dd x \\
&=&\int_{0}^{\infty}\!\!\int_{0}^{\infty}K_{00}(x,v)W(x,v)\dd v\dd x=m \\
\int_{0}^{\infty}K_{10}(x,x)\dd x&=&\int_{0}^{\infty}\!\!\int_{0}^{\infty}x^{\alpha+1}e^{-x}\frac{w^{\alpha}\e^{-w}}{x+w}K_{00}(w,x)\dd w\dd x \\
&=&\int_{0}^{\infty}\!\!\int_{0}^{\infty}K_{00}(w,x)W(w,x)\dd w\dd x=m.
\end{eqnarray}

To verify the normalization constant $4m(m-1)$ of the two-point density~(\ref{eq:h2}), we first note that
\begin{equation}
\int_{0}^{\infty}\!\!\int_{0}^{\infty}(\left(K_{01}(x,x)+K_{10}(x,x)\right)\left(K_{01}(y,y)+K_{10}(y,y)\right)\dd x\dd y=4m^{2}.
\end{equation}
We now compute
\begin{eqnarray}
&&\int_{0}^{\infty}\!\!\int_{0}^{\infty}K_{01}(x,y)K_{01}(y,x)\dd x\dd y \\
&=&\int_{0}^{\infty}\!\!\int_{0}^{\infty}\!\!\int_{0}^{\infty}\!\!\int_{0}^{\infty}x^{\alpha}e^{-x}\frac{v^{\alpha+1}\e^{-v}}{x+v}K_{00}(y,v)y^{\alpha}e^{-y}\frac{w^{\alpha+1}\e^{-w}}{y+w}K_{00}(x,w)\dd v\dd w\dd x\dd y \nonumber\\
&=&\int_{0}^{\infty}\!\!\int_{0}^{\infty}K_{00}(y,w)y^{\alpha}e^{-y}\frac{w^{\alpha+1}\e^{-w}}{y+w}\dd w\dd y=m.
\end{eqnarray}
Similarly, we have
\begin{eqnarray}
&&\int_{0}^{\infty}\!\!\int_{0}^{\infty}K_{10}(x,y)K_{10}(y,x)\dd x\dd y \\
&=&\int_{0}^{\infty}\!\!\int_{0}^{\infty}\!\!\int_{0}^{\infty}\!\!\int_{0}^{\infty}y^{\alpha+1}e^{-y}\frac{w^{\alpha}\e^{-w}}{y+w}K_{00}(w,x)x^{\alpha+1}e^{-x}\frac{v^{\alpha}\e^{-v}}{x+v}K_{00}(v,y)\dd w\dd v\dd x\dd y \nonumber\\
&=&\int_{0}^{\infty}\!\!\int_{0}^{\infty}y^{\alpha+1}e^{-y}\frac{w^{\alpha}\e^{-w}}{y+w}K_{00}(w,y)\dd w\dd y=m.
\end{eqnarray}
Finally, one computes
\begin{eqnarray*}
&&\int_{0}^{\infty}\!\!\int_{0}^{\infty}K_{00}(x,y)K_{11}(x,y)\dd x\dd y\left(=\int_{0}^{\infty}\!\!\int_{0}^{\infty}K_{00}(y,x)K_{11}(y,x)\dd x\dd y\right)\nonumber\\
&=&\int_{0}^{\infty}\!\!\int_{0}^{\infty}K_{00}(x,y)\left(x^{\alpha}y^{\alpha+1}e^{-x-y}\int_{0}^{\infty}\!\!\int_{0}^{\infty}\frac{v^{\alpha}\e^{-v}}{y+v}\frac{w^{\alpha+1}\e^{-w}}{x+w}K_{00}(v,w)\dd v\dd w-W(x,y)\right)\dd x\dd y \nonumber\\
&=&\sum_{i,j=0}^{m-1}\int_{0}^{\infty}\!\!\int_{0}^{\infty}\widetilde{p}_{i}(x)\widetilde{q}_{j}(w)\frac{x^{\alpha}w^{\alpha+1}\e^{-x-w}}{x+w}\dd x\dd w\int_{0}^{\infty}\!\!\int_{0}^{\infty}\widetilde{p}_{j}(v)\widetilde{q}_{i}(y)\frac{v^{\alpha}y^{\alpha+1}\e^{-v-y}}{v+y}\dd v\dd y-m \nonumber\\
&=&\sum_{i=j=0}^{m-1}1-m=0,
\end{eqnarray*}
where for convenience we used the non-monic version~\cite{bertola14} of the biorthogonal polynomials $\widetilde{p}_{k}(x)$ and $\widetilde{q}_{l}(y)$, i.e.,
\begin{equation}
K_{00}(x,y)=\sum_{k=0}^{m-1}\widetilde{p}_{k}(x)\widetilde{q}_{k}(y)
\end{equation}
and
\begin{equation}
\int_{0}^{\infty}\!\!\int_{0}^{\infty}\widetilde{p}_{k}(x)\widetilde{q}_{l}(y)W(x,y)\dd x\dd y=\delta_{kl}.
\end{equation}
We now arrive at the normalization constant
\begin{equation}
4m^{2}-2m-2m-0=4m(m-1).
\end{equation}

Before discussing the $3$-point density normalization, we compute two additional integral identities. The first one is
\begin{eqnarray}
&&\int_{0}^{\infty}\!\!\int_{0}^{\infty}K_{01}(x,y)K_{10}(x,y)\dd x\dd y\left(=\int_{0}^{\infty}\!\!\int_{0}^{\infty}K_{01}(y,x)K_{10}(y,x)\dd x\dd y\right)\nonumber\\
&=&\int_{0}^{\infty}\!\!\int_{0}^{\infty}\!\!\int_{0}^{\infty}\!\!\int_{0}^{\infty}x^{\alpha}\e^{-x}\frac{v^{\alpha+1}\e^{-v}}{x+v}K_{00}(y,v)y^{\alpha+1}\e^{-y}\frac{w^{\alpha}\e^{-w}}{y+w}K_{00}(w,x)\dd v\dd w\dd x\dd y\nonumber\\
&=&\sum_{i,j=0}^{m-1}\int_{0}^{\infty}\!\!\int_{0}^{\infty}\!\!\int_{0}^{\infty}\!\!\int_{0}^{\infty}x^{\alpha}\e^{-x}\frac{v^{\alpha+1}\e^{-v}}{x+v}\widetilde{p}_{i}(y)\widetilde{q}_{i}(v)y^{\alpha+1}\e^{-y}\frac{w^{\alpha}\e^{-w}}{y+w}\widetilde{p}_{j}(w)\widetilde{q}_{j}(x)\dd v\dd w\dd x\dd y\nonumber\\
&=&\sum_{i,j=0}^{m-1}\int_{0}^{\infty}\!\!\int_{0}^{\infty}\frac{x^{\alpha}v^{\alpha+1}\e^{-x-v}}{x+v}\widetilde{q}_{j}(x)\widetilde{q}_{i}(v)\dd x\dd v\int_{0}^{\infty}\!\!\int_{0}^{\infty}\frac{w^{\alpha}y^{\alpha+1}\e^{-w-y}}{w+y}\widetilde{p}_{j}(w)\widetilde{p}_{i}(y)\dd w\dd y\nonumber\\
&=&\sum_{i,j=0}^{m-1}\int_{0}^{\infty}\!\!\int_{0}^{\infty}\frac{x^{\alpha}v^{\alpha+1}\e^{-x-v}}{x+v}(\widetilde{p}_{j}(x)+\text{l.o.t.})\widetilde{q}_{i}(v)\dd x\dd v\nonumber\\
&&\times\int_{0}^{\infty}\!\!\int_{0}^{\infty}\frac{w^{\alpha}y^{\alpha+1}\e^{-w-y}}{w+y}\widetilde{p}_{j}(w)(\widetilde{q}_{i}(y)+\text{l.o.t.})\dd w\dd y\nonumber\\
&=&\sum_{i<j=0}^{m-1}\text{(non-zero contribution)}\times 0+\sum_{i>j=0}^{m-1}0\times\text{(non-zero contribution)}+\sum_{i=j=0}^{m-1}1\nonumber\\
&=&m,\label{eq:0110}
\end{eqnarray}
where we have used the fact that the coefficients of the highest degree terms of the polynomials $\widetilde{p}_{j}(x)$ and $\widetilde{q}_{j}(y)$ are the same~\cite{bertola14}. By the relations~(\ref{eq:vw}), the correlation kernels satisfy~\cite{bertola14}
\begin{eqnarray*}
&&K_{01}(x,y)K_{01}(y,x)+K_{10}(x,y)K_{10}(y,x)+K_{00}(x,y)K_{11}(x,y)+K_{00}(y,x)K_{11}(y,x)\nonumber \\
&=&K_{01}(x,y)K_{10}(x,y)+K_{01}(y,x)K_{10}(y,x)-K_{00}(x,y)K_{11}(y,x)-K_{00}(y,x)K_{11}(x,y),
\end{eqnarray*}
which, with the previously obtained results, leads to the identities
\begin{equation}
\int_{0}^{\infty}\!\!\int_{0}^{\infty}K_{00}(x,y)K_{11}(y,x)\dd x\dd y=\int_{0}^{\infty}\!\!\int_{0}^{\infty}K_{00}(y,x)K_{11}(x,y)\dd x\dd y=0.
\end{equation}
To verify the normalization constant $8m(m-1)(m-2)$ of the three-point density~(\ref{eq:h3}), we start from the integrals over the terms~(\ref{eq:hA}) and~(\ref{eq:hB}), which are given by $(2m)^{3}=8m^{3}$ and $3\times(-2\times2m\times2m)=-24m^{2}$, respectively. To compute integrals of the terms~(\ref{eq:hC}), we first compute some intermediate results. One has
\begin{eqnarray}
&&\int_{0}^{\infty}K_{00}(x,y)K_{11}(z,y)\dd y\nonumber\\
&=&\int_{0}^{\infty}K_{00}(x,y)\left(z^{\alpha}y^{\alpha+1}e^{-z-y}\int_{0}^{\infty}\!\!\int_{0}^{\infty}\frac{v^{\alpha}\e^{-v}}{y+v}\frac{w^{\alpha+1}\e^{-w}}{z+w}K_{00}(v,w)\dd v\dd w-W(z,y)\right)\dd y\nonumber\\
&=&\int_{0}^{\infty}\!\!\int_{0}^{\infty}\frac{z^{\alpha}w^{\alpha+1}\e^{-z-w}}{z+w}K_{01}(v,x)K_{00}(v,w)\dd v\dd w-\int_{0}^{\infty}K_{00}(x,y)W(z,y)\dd y\nonumber\\
&=&\int_{0}^{\infty}K_{01}(z,v)K_{01}(v,x)\dd v-K_{01}(z,x)=K_{01}(z,x)-K_{01}(z,x)=0
\end{eqnarray}
and
\begin{eqnarray}
&&\int_{0}^{\infty}K_{00}(x,y)K_{11}(x,z)\dd x\nonumber\\
&=&\int_{0}^{\infty}K_{00}(x,y)\left(x^{\alpha}z^{\alpha+1}e^{-x-z}\int_{0}^{\infty}\!\!\int_{0}^{\infty}\frac{v^{\alpha}\e^{-v}}{z+v}\frac{w^{\alpha+1}\e^{-w}}{x+w}K_{00}(v,w)\dd v\dd w-W(x,z)\right)\dd x\nonumber\\
&=&\int_{0}^{\infty}\!\!\int_{0}^{\infty}K_{00}(v,w)\frac{v^{\alpha}z^{\alpha+1}\e^{-v-z}}{v+z}\dd v\int_{0}^{\infty}K_{00}(x,y)\frac{x^{\alpha}w^{\alpha+1}\e^{-x-w}}{x+w}\dd x\dd w-\nonumber\\
&&\int_{0}^{\infty}K_{00}(x,y)W(x,z)\dd x\nonumber\\
&=&\int_{0}^{\infty}K_{10}(w,z)K_{10}(y,w)\dd w-K_{10}(y,z)=K_{10}(y,z)-K_{10}(y,z)=0,
\end{eqnarray}
which respectively lead to
\begin{equation}
0=\int_{0}^{\infty}K_{00}(x,y)K_{11}(z,y)\dd y\int_{0}^{\infty}\frac{x^{\alpha}v^{\alpha+1}\e^{-x-v}}{x+v}\dd x=\int_{0}^{\infty}K_{10}(y,v)K_{11}(z,y)\dd y
\end{equation}
and
\begin{equation}
0=\int_{0}^{\infty}K_{00}(x,y)K_{11}(x,z)\dd x\int_{0}^{\infty}\frac{v^{\alpha}y^{\alpha+1}\e^{-v-y}}{v+y}\dd y=\int_{0}^{\infty}K_{01}(v,x)K_{11}(x,z)\dd x.
\end{equation}
One also has
\begin{eqnarray}
&&\int_{0}^{\infty}K_{00}(x,y)K_{01}(x,z)\dd x\nonumber\\
&=&\int_{0}^{\infty}\!\!\int_{0}^{\infty}K_{00}(x,y)\frac{x^{\alpha}v^{\alpha+1}\e^{-x-v}}{x+v}K_{00}(z,v)\dd x\dd v\nonumber\\
&=&\sum_{i,j=0}^{m-1}\widetilde{p}_{j}(z)\widetilde{q}_{i}(y)\int_{0}^{\infty}\!\!\int_{0}^{\infty}\widetilde{p}_{i}(x)\widetilde{q}_{j}(v)\frac{x^{\alpha}v^{\alpha+1}\e^{-x-v}}{x+v}\dd x\dd v\nonumber\\
&=&\sum_{i=0}^{m-1}\widetilde{p}_{i}(z)\widetilde{q}_{i}(y)=K_{00}(z,y)
\end{eqnarray}
and
\begin{eqnarray}
&&\int_{0}^{\infty}K_{00}(x,y)K_{10}(z,y)\dd y\nonumber\\
&=&\int_{0}^{\infty}\!\!\int_{0}^{\infty}K_{00}(x,y)\frac{w^{\alpha}y^{\alpha+1}\e^{-w-y}}{w+y}K_{00}(w,z)\dd w\dd y\nonumber\\
&=&\sum_{i,j=0}^{m-1}\widetilde{p}_{i}(x)\widetilde{q}_{j}(z)\int_{0}^{\infty}\!\!\int_{0}^{\infty}\widetilde{p}_{j}(w)\widetilde{q}_{i}(y)\frac{w^{\alpha}y^{\alpha+1}\e^{-w-y}}{w+y}\dd w\dd y\nonumber\\
&=&\sum_{i=0}^{m-1}\widetilde{p}_{i}(x)\widetilde{q}_{i}(z)=K_{00}(x,z).
\end{eqnarray}
With the above integral identities, it can be verified that each of the $24$ terms in~(\ref{eq:hC}) has a zero contribution to the normalization. At the same time, each of the $8$ terms in~(\ref{eq:hD}) integrates to $m$ by using the reproducing properties~(\ref{eq:re01}) and~(\ref{eq:re10}) as well as the identity~(\ref{eq:0110}). The normalization constant is therefore obtained as
\begin{equation}
8m^{3}-24m^{2}+2\times 8m=8m(m-1)(m-2).
\end{equation}

\end{document}